\documentclass[mnsc,nonblindrev]{informs3_hide} 

\OneAndAHalfSpacedXI 

\usepackage{natbib,multirow,multicol,xspace,enumitem,subcaption,caption}
\usepackage{accents}
\bibpunct[, ]{(}{)}{,}{a}{}{,}%
\def\bibfont{\small}%
\usepackage{bm}
\usepackage[normalem]{ulem}
\usepackage[dvipsnames]{xcolor}
\usepackage[colorlinks=true,breaklinks=true,bookmarks=false,urlcolor=blue,citecolor=blue,linkcolor=blue,bookmarksopen=false,draft=false]{hyperref}
\hypersetup{hidelinks}

\newcommand{\eps}{\varepsilon}

\newcommand{\bR}{\mathbb{R}}
\newcommand{\bE}{\mathbb{E}}

\newcommand{\cS}{\mathcal{S}}
\newcommand{\OPT}{\mathsf{OPT}}
\newcommand{\OPTp}{\mathsf{OPT}^+}

\newcommand{\ALG}{\mathsf{ALG}}
\newcommand{\THR}{\mathsf{THR}}
\newcommand{\Gre}{\mathsf{Greedy}}
\newcommand{\size}{\mathsf{size}}
\newcommand{\ALGN}{\mathsf{ALG}_{\textit{\ref{defn:3/7alg}}}}
\newcommand{\FN}{F_{\textit{\ref{defn:3/7alg}}}}

\newcommand{\ALGG}{\mathsf{ALG}_{\textit{\ref{defn:0.4324alg}}}}
\newcommand{\FG}{F_{\textit{\ref{defn:0.4324alg}}}}
\newcommand{\cG}{c_{\textit{\ref{defn:0.4324alg}}}}
\newcommand{\qG}{q_{\textit{\ref{defn:0.4324alg}}}}

\usepackage{color}              
\usepackage{color-edits}
\addauthor{Will}{red}

\TheoremsNumberedThrough     
\ECRepeatTheorems

\EquationsNumberedThrough    

\MANUSCRIPTNO{}

\begin{document}



\RUNTITLE{Threshold Policies for Online Knapsack}

\TITLE{The Competitive Ratio of Threshold Policies for Online Unit-density Knapsack Problems}

\ARTICLEAUTHORS{
\AUTHOR{Will Ma}
\AFF{Graduate School of Business, Columbia University, New York, NY 10027, \EMAIL{wm2428@gsb.columbia.edu}}
\AUTHOR{David Simchi-Levi}
\AFF{Institute for Data, Systems, and Society, Department of Civil and Environmental Engineering, and Operations Research Center, Massachusetts Institute of Technology, Cambridge, MA 02139, \EMAIL{dslevi@mit.edu}}
\AUTHOR{Jinglong Zhao}
\AFF{Questrom School of Business, Boston University, Boston, MA 02215, \EMAIL{jinglong@bu.edu}}
}

\ABSTRACT{
We study a wholesale supply chain ordering problem. In this problem, the supplier has an initial stock, and faces an unpredictable stream of incoming orders, making real-time decisions on whether to accept or reject each order. What makes this wholesale supply chain ordering problem special is its ``knapsack constraint,'' that is, we do not allow partially accepting an order or splitting an order. The objective is to maximize the utilized stock.

We model this wholesale supply chain ordering problem as an online unit-density knapsack problem. We study randomized threshold algorithms that accept an item as long as its size exceeds the threshold. We derive two optimal threshold distributions, the first is 0.4324-competitive relative to the optimal offline integral packing, and the second is 0.4285-competitive relative to the optimal offline fractional packing. Both results require optimizing the cumulative distribution function of the random threshold, which are challenging infinite-dimensional optimization problems. We also consider the generalization to multiple knapsacks, where an arriving item has a different size in each knapsack. We derive a 0.2142-competitive algorithm for this problem. We also show that any randomized algorithm for this problem cannot be more than 0.4605-competitive. This is the first upper bound strictly less than 0.5, which implies the intrinsic challenge of knapsack constraint.

We show how to naturally implement our optimal threshold distributions in the warehouses of a Latin American chain department store. We run simulations on their order data, which demonstrate the efficacy of our proposed algorithms.
}

\vspace{-2cm}

\maketitle

\section{Introduction}

This paper considers a wholesale supply chain ordering problem faced by a wholesale supplier managing a single product.
The single product has different specifications, each associated with an initial stock level. 
The supplier faces an unpredictable stream of incoming orders, making real-time decisions on whether to accept or reject each order.
An order can only be accepted if its size does not exceed the remaining stock of the chosen specification, and cannot be split across multiple specifications.
The objective is to maximize the sum of sizes of accepted orders, that is, to maximize the total utilized stock.


This wholesale supply chain ordering problem differs from traditional supply chain ordering problems or online resource allocation problems such as the AdWords problem \citep{mehta2005adwords, huang2020adwords}.
One major distinction is the ``knapsack constraint,'' that is, we do not allow partially accepting an order or splitting an order across multiple specifications.
As \citet{simchi2005logic} noted, this is because ``splitting an order is often physically impossible or managerially undesirable due to customer service or accounting considerations.''
This paper thus studies the new challenges that the knapsack constraint brings to wholesale supply chains.

In Examples~\ref{exa:CDS} and~\ref{exa:Cocoa} below, we illustrate the wholesale supply chain ordering problem and illustrate the knapsack constraint in specific supply chain contexts.
We will provide more examples in Section~\ref{sec:Literature} to illustrate variants of this problem, and review other related problems.

\begin{example}[Chain Department Store]
\label{exa:CDS}
A Latin American chain department store manages 974 Stock Keeping Units (SKUs) in the young women's fashion category. 
The chain operates 21 regional warehouses, each serving a specific set of local stores in that region, with each store being served by only one warehouse. 
We manage each SKU-warehouse pair independently. 
Each order from a local store cannot be split or redirected to a different warehouse because of the long shipment time and the high managerial costs involved. 
For more details on this chain department store, see Section~\ref{sec::computational}.
\end{example}

\begin{example}[Cocoa distribution]
\label{exa:Cocoa}
A multinational food and beverage manufacturing company distributes cocoa to its downstream retailers.
This company sources cocoa from three West African countries, which we refer to as three distinct specifications. 
Variations in the growing regions result in different proportions of cocoa solids and cocoa butter. 
Due to the differences in cocoa composition, we allow each order to request different amounts under different specifications; but mixing specifications is disallowed to ensure quality and consistency. 
For details on how to capture this example using our model, see Example~\ref{exa:CocoaContinued} in Section~\ref{sec::AdWords}.
\end{example}

As illustrated in Examples~\ref{exa:CDS} and~\ref{exa:Cocoa}, the wholesale supply chain ordering problem has two variations.
Example~\ref{exa:CDS} illustrates a ``single-knapsack'' problem, where each SKU-warehouse pair is modeled as a single knapsack and the department store manages all the knapsacks independently.
In contrast, Example~\ref{exa:Cocoa} illustrates a ``multi-knapsack'' problem, where each specification of cocoa is modeled as a knapsack and the company manages three knapsacks jointly.
The single-knapsack problem can be seen as a special case of the multi-knapsack problem, with solutions to the single-knapsack problem serving as key ingredients for solving the multi-knapsack problem.

Below we examine the single-knapsack problem and introduce the motivation for studying threshold policies.
To make our terminology consistent, we will use ``item'' to represent an order, ``stock'' to represent capacity, and ``knapsack'' to represent a specification (in the multi-knapsack problem).



\subsection{Single-Knapsack Problem and Threshold Policies}
\label{sec:SingleKnapsackThresholdPolicy}

In the single-knapsack problem, we can normalize the size of the knapsack to be 1.
In this problem the supplier's decision is simply whether to accept or reject an incoming item, and there is no routing decision to make. 
This decision must be made without knowing the sizes or the number of future items, nor the number of future items.
We refer to this decision rule as an ``online algorithm.'' 
Conversely, if the entire sequence of items were known in advance, the supplier could optimally choose which items to accept or reject.
We refer to this decision rule as an ``offline algorithm.'' 
An (possibly randomized) online algorithm is said to be $c$-competitive for any constant $c \leq 1$ if, for any sequence of items, its (expected) packed capacity is at least $c$ times that of an optimal offline algorithm. 
We are interested in finding the highest possible value of $c$, which is called the ``competitive ratio.''


In the single-knapsack problem, \citet{han2015randomized} already proposed a $\frac{1}{2}$-competitive algorithm.
First, a fair coin is flipped and generates two scenarios: Head and Tail.
If Head, the algorithm greedily packs any item that fits.
If Tail, the algorithm rejects all items until the first one that the greedy policy would not have fit, and starts to greedily accept from that item.
In expectation, this algorithm is $\frac{1}{2}$-competitive, since either the total size of all items is no more than 1, in which case greedy (which the algorithm mimics with half probability) is optimal; or the total size of all items is more than 1, in which case the algorithm's sum of capacity packed under the two scenarios exceeds 1 and hence its expected packing exceeds $\frac{1}{2}$, while the offline optimum is at most 1.
Furthermore, it is easy to show that no online algorithm performs better than $\frac{1}{2}$-competitive.

Although the algorithm in \citet{han2015randomized} already achieves the best possible competitive ratio on a single knapsack (yet \citet{han2015randomized} does not study multi-knapsacks), this algorithm is not simple enough to communicate to supply chain managers.
Specifically, if Tail, the algorithm proposed in \citet{han2015randomized} requires rejecting the first few arriving items until the first one that the greedy policy would not have fit.
Our industry partner, the Latin American chain department store mentioned in Example~\ref{exa:CDS}, finds it difficult to explain to their early-arriving customers (i.e., local stores) why their orders are rejected while later orders are accepted.
See Example~\ref{exa:Complaints} for an illustration.

\begin{example}[Customer Complaints]
\label{exa:Complaints}
Consider a single-knapsack problem with a sequence of items whose sizes are $(0.4, 0.5, 0.5)$. 
The algorithm in \citet{han2015randomized}, if the scenario is Head, will accept the first two items, $0.4$ and $0.5$. 
If the scenario is Tail, the algorithm will reject the first two items and accept only the last item, $0.5$. 
In the Tail scenario, even though the second and third items are the same size, and the remaining capacity is identical when both items arrive, the second item, which arrives earlier, is rejected.
\end{example}

Example~\ref{exa:Complaints} illustrates the challenge of applying the algorithm in \citet{han2015randomized} with our industry partner.
One type of algorithms that they find explainable in their setting is the family of randomized threshold algorithms.
A randomized threshold algorithm initially draws a random threshold $\tau$ that takes values between $[0,1]$, and then accepts every item of size at least $\tau$ that fits, but never changes the threshold throughout the entire horizon after its initial draw.
When $\tau=0$, the algorithm mimics the greedy algorithm.

\begin{example}[Example~\ref{exa:Complaints} Continued]
\label{exa:NoComplaints}
We consider the same single-knapsack problem and the same sequence of items whose sizes are $(0.4, 0.5, 0.5)$. 
A threshold algorithm with a threshold $\tau > 0.5$ will reject both the second and third items. 
A threshold algorithm with a threshold $\tau \leq 0.5$ will accept the second and third items as long as there is remaining capacity.
It is possible that the second item is accepted when there is sufficient remaining capacity, yet the third item is rejected when there is insufficient remaining capacity.
Our industry partner can easily explain by pointing out that the third item arrives later.
\end{example}

Example~\ref{exa:NoComplaints} illustrates the ``explainability'' of threshold policies in our industry partner's supply chain. 
Beyond their explainability, threshold policies offer an additional benefit of ``simplicity.'' 
These policies are logistically simple to implement in a single-knapsack setting, allowing for easy accept or reject decisions based solely on the remaining capacity, without the need to track the history of past items.
Although the benefits of explainability and simplicity are difficult to quantify mathematically, we can quantify the cost of restricting ourselves to threshold policies by a small gap in competitive ratios.
Our results support industry practitioners to prioritize the practical advantages of adopting threshold policies over their associated costs.
See Section~\ref{sec::introGeneralization} below for further details.

Finally, we provide a different model of selling services in which threshold policies can be implemented but the policy of \cite{han2015randomized} cannot. 
This model is reasonable whenever there is a market-accepted price per unit time of the service, and customers want a fixed amount. 
Some plausible applications are car/parking rentals, hotel bookings, and cloud computing services.

\begin{example}[Revenue Maximization] \label{eg:RM}
A service provider is selling reservations to get service for a fixed amount of time, before the service time window begins. 
The time window is normalized to have length 1. 
All customers $t$ have the same, known willingness-to-pay rate $\rho$ for this service, but want different, fixed amounts $s_t$.  
That is, the valuation function of customer $t$ is $V_t(x)=\begin{cases} \rho s_t, & x\ge s_t\\ 0, & x< s_t \\ \end{cases}$, where $x$ is the amount they receive.  The customers arrive one-by-one and the sequence $(s_1,\ldots,s_T)$ is unknown and may not even be truthfully revealed upon arrival.  The goal is to set a pricing scheme to maximize revenue.
\end{example}

In Example~\ref{eg:RM}, a threshold of $\tau$ would correspond to setting a price of $\rho\cdot\max\{x,\tau\}$ to use any amount $x\ge0$ that is less than the remaining amount.  Customers $t$ with $s_t<\tau$ would not pay for the service (because they would have to pay a minimum of $\rho\tau$, which is greater than their maximum possible valuation of $\rho s_t$), whereas those with $s_t\ge\tau$ would pay $\rho s_t$ to use $x=s_t$ as long as the remaining amount is at least $s_t$.  The total revenue collected would be exactly $\rho$ times the sum of usage of paying customers.  By contrast, the policy of \citet{han2015randomized} cannot be implemented because it requires knowing the sizes of customers that do not use the service.

\subsection{Generalization to Multi-Knapsack Problem and Summary of Results}
\label{sec::introGeneralization}

On a single knapsack, we derive tight competitive ratios for randomized threshold algorithms under two different definitions of the optimal offline algorithm: a $3/7\approx0.4285$-competitive algorithm relative to the optimal offline ``fractional'' packing (\textbf{Theorems~\ref{thm:3/7} and~\ref{thm:3/7example}}); and a $0.4324$-competitive algorithm relative to the optimal offline ``integer'' packing (\textbf{Theorems~\ref{thm:0.4324} and~\ref{thm:0.4324example}}).
The optimal offline fractional packing allows partially accepting an item, while the optimal offline integer packing requires accepting an item in its entirety.
The integer packing is a reasonable comparison because any threshold algorithm must also accept an item in its entirety.
However, the fractional packing becomes more relevant when we generalize to the multi-knapsack problem.

We generalize our single knapsack results to multiple knapsacks by proposing a general recipe that combines routing algorithms (which determine the knapsack to which an item is routed) with accept-or-reject algorithms (which decide whether to accept an item on a single knapsack).
This recipe implies a $3/14\approx0.2142$-competitive algorithm (\textbf{Theorem~\ref{thm:dependentSize}}) by combining a greedy routing policy that is well-studied in the literature \citep{mehta2013online}, and the randomized threshold policy as we designed for the single-knapsack problem.
Under this combined policy, each item is first routed to the knapsack where it could potentially take the largest size, without knowing whether it will be accepted.
Then an independent threshold policy at that knapsack controls whether to accept the item.
To implement this routing policy, we need to introduce a ``phantom'' process in which all routed items are accepted, and route each incoming item to the knapsack in which it would take the greatest size after truncated by the remaining phantom capacity.

Such a recipe is general and can be used to combine other routing algorithms with other accept-or-reject algorithms.
Recent developments on the AdWords problem, such as the $0.5016$-competitive algorithm in \citet{huang2020adwords}, can be used to replace the aforementioned greedy routing algorithm.
The algorithm proposed in \citet{han2015randomized} can also be used to replace the threshold algorithm.
Implementing these replacements can straightforwardly improve the $\frac{3}{14}\approx0.2142$-competitive algorithm to $0.2508$-competitive, as we explain at the end of Section~\ref{sec:multipleALG}.
However, we cannot use our $0.4324$-competitive algorithm for multiple knapsacks in this way, because our recipe for combination requires the competitive ratio to hold relative to an optimal offline fractional packing.

For multiple knapsacks we derive an upper bound of $0.4605$ on the competitive ratio for arbitrary randomized algorithms (\textbf{Theorem~\ref{thm:knapsackDEexample}}).
This indicates that the tight $\frac{1}{2}$-competitiveness for a single knapsack does not extend to multiple knapsacks, even if one could go beyond threshold policies.
In comparison, the state-of-the-art AdWords algorithm from \citet{huang2020adwords}, which allows for partial acceptance of items, achieves a competitive ratio of $0.5016$ that is greater than our upper bound of $0.4605$. 
This comparison illustrates the fundamental challenge that the knapsack constraint introduces to the wholesale supply chain ordering problem, as we have discussed in Examples~\ref{exa:Cocoa} and~\ref{exa:CDS}.
We summarize all our aforementioned results in Table~\ref{tbl:summaryResults}.

\begin{table}[!tb]
\TABLE
{Summary of lower and upper bounds on the competitive ratios.
\label{tbl:summaryResults}}
{
\begin{tabular}{|l|c|c|}
\hline
& Relative to Stronger Optimum & Relative to Weaker Optimum \\
\hline
\updown Single Knapsack
& [0.4285 (\textbf{Thm.~\ref{thm:3/7}}), 0.4285 (\textbf{Thm.~\ref{thm:3/7example}})]
& [0.4324 (\textbf{Thm.~\ref{thm:0.4324}}), 0.4324 (\textbf{Thm.~\ref{thm:0.4324example}})] \\
\hline
\updown Multiple Knapsacks
& [0.2142 (\textbf{Thm.~\ref{thm:dependentSize}}), $\longrightarrow$] & [$\longleftarrow$, 0.4605 (\textbf{Thm.~\ref{thm:knapsackDEexample}})] \\
\hline
\updown AdWords (truncation allowed) & [0.5016 \citep{huang2020adwords}, $\longrightarrow$] & [$\longleftarrow$, 0.6321 \citep{karp1990optimal}] \\
\hline
\updown Scheduling (stochastic arrivals) & [0.355 \citep{jiang2024tight}, $\longrightarrow$] & [$\longleftarrow$, 0.5 \citep{stein2018advance}] \\
\hline
\end{tabular}
}
{This table presents lower and upper bounds on the competitive ratios for different classes of algorithms, relative to different optima, in different settings. 
Results from this paper are bolded. 
An arrow indicates that the best-known lower (resp.\ upper) bound is implied by that from a more restricted (resp.\ less restricted) setting, pointing in the direction of that setting. 
Note that our paper is the only one to establish a separation between the competitiveness relative to the two different optima.
Note also that for the AdWords problem, the 0.5016 result in \citet{huang2020adwords} improves to $1-\frac{1}{e}\approx0.6321$ under the ``small bids'' assumption \citep{mehta2005adwords}.}
\end{table}

\subsection{Related Works and Applications}
\label{sec:Literature}

To the best of our knowledge, we are the first to study threshold policies on the single-knapsack problem and derive tight constant-factor competitiveness results.
We also consider the unit-density multi-knapsack problem that can be viewed as a harder version of Adwords, and derive both constant-factor competitiveness and impossibility results.

There is a rich literature which studies online knapsack problems, with one of the earliest works dating back to \cite{marchetti1995stochastic}.
In the general online knapsack problem, each item is associated with a size and a reward.
The unit-density knapsack problem can be seen as a special case of the general problem when the reward is equal (or proportional) to the size.
Without the unit-density assumption, the non-existence of any constant competitive ratio guarantee, even for randomized algorithms on a single knapsack, was first established in \cite{marchetti1995stochastic}.
Tight instance-dependent competitive ratios (where the competitive ratio can depend on parameters based on the sequence of items) have also been established in \citet{zhou2008budget}.
For a thorough discussion of technical results across many variants of online knapsack, we refer to \citet{cygan2016online}.
More recent references can be found in the papers \citet{sun2022online,lechowicz2023time,daneshvaramoli2024competitive}, which consider new directions for the online knapsack problem that include multiple knapsacks, data-driven algorithm tuning, time fairness, and online knapsack with predictions.

There is also a rich literature which studies online matching problems, with the earliest work dating back to \citet{karp1990optimal}. 
In this problem, there is a bipartite graph where one side of the nodes (RHS) arrive sequentially. Upon the arrival of one node, the decision maker has to make an irrevocable decision that matches this RHS node to another on the other side (LHS). The problem is to maximize the total number of matched pairs. 
In this problem, there exists a greedy algorithm that achieves a $1/2$-competitive ratio. 
\citet{karp1990optimal} proposed a randomized algorithm called $\mathsf{RANKING}$ that achieves a $1-\frac{1}{e}$-competitive ratio.
This problem was first generalized to the online $b$-matching problem, a more general setting where each node on the left side of the graph can be matched to no more than $b$ different RHS nodes \citet{kalyanasundaram1993online}. 

The online knapsack problem and the online matching problem cross at the the famous AdWords problem \citep[see, e.g.,][]{devanur2009AdWords, devanur2012online, goel2008online, huang2020adwords, mehta2005adwords}.
The AdWords problem can be seen as an online knapsack problem under the unit-density assumption and that allows for partial acceptance.
The AdWords problem can also be seen as a generalized online matching problem where each node on the left side has a budget to allocate to the right side nodes. 
This problem is motivated from online advertising, where, as noted in \citep{mehta2005adwords}, ``a search engine company decide what advertisements to display with each query so as to maximize its revenue.'' 
Consumers arrive in an online fashion, type keywords, and reveal their preferences. 
A search engine then displays personalized advertisements to the consumer, and earns monetary transfers from companies who bid for consumer keywords within their daily budgets. 
The objective is to maximize the total revenue earned from the companies.

There is a greedy algorithm, which we will describe in Definition~\ref{defn:AdWordsGreedy}, that is $\frac{1}{2}$-competitive for the AdWords problem.
Further assuming the ``small bids'' assumption that the sizes of the items are much smaller than the initial capacities of the knapsacks, \citet{mehta2005adwords} showed that an algorithm called $\mathsf{BALANCE}$ achieves an $1-\frac{1}{e} \approx 0.6321$ competitive ratio.
Our problem reduces to the AdWords problem under the small bid assumption, because allowing for partial acceptance does not make a difference when the items are small.

Other than the online knapsack and online matching problems described above, there is also a parallel literature which studies the stochastic variant of the online knapsack and online matching problems.
This literature assumes the arrival sequences are drawn from a given distribution. 
There are papers on the optimal policies on a single knapsack; see \citet{kleywegt1998dynamic, papastavrou1996dynamic} when the order of arriving items is fixed.
When the items can be inserted in any order but their sizes are stochastic, the concept of ``adaptivity gap'' between adaptive and non-adaptive algorithms was proposed in \citet{dean2008approximating}.
Variants where the arrival sequence could be learned over time are studied in \citet{modaresi2019learning, hwang2018online}.
Under the unit-density assumption, \citet{stein2018advance} has provided competitive ratio type of guarantees. 
They show a $0.321$ ratio against the expected performance of the optimal decision maker's expected revenue, which has been recently improved to 0.355 by \citet{jiang2024tight}. 
\citet{jiang2024tight} also establish a smaller competitive ratio of $1/(3+e^{-2})\approx0.319$ for the stochastic setting without the unit-density assumption.
We summarize the aforementioned related results in Table~\ref{tbl:summaryResults}.

Finally, our adherence to threshold policies is motivated by the importance of simplicity in pricing and mechanism design---see \citet{wang2023power} for a recent reference.  As discussed in Example~\ref{eg:RM}, the benefit of online algorithms defined by thresholds is that they are incentive-compatible, which can be useful even for welfare maximization \citep{feldman2014combinatorial}.  For further discussions about the virtues of threshold policies, we refer to \citet{arnosti2023tight,elmachtoub2023power}.




\subsection{Simulations Using Data from a Latin American Chain Department Store} \label{sec::introSimulations}

We now describe how our optimal random-threshold policy for a single knapsack can be implemented across the supply chain of our industry partner, a Latin American chain department store.
They sell 974 Store Keeping Units (SKUs) in the young women's fashion category.
There are 21 warehouses, and every SKU is stored in a subset of different warehouses.
Each $(\text{SKU}, \text{warehouse})$-pair faces a stream of orders, each for a specific number of units.
Orders cannot be split or redirected to a different warehouse, as there are only a few warehouses across the entire country, and our industry partner is unwilling to bear the long shipment time or the high managerial cost.
Therefore, our industry partner faces the same accept/reject problem on order sizes, and has the same goal of maximizing total inventory fulfilled.

Our industry partner is specially in favor of simple threshold policies, as such policies are believed to be more relevant in practice.
See Examples~\ref{exa:Complaints} and~\ref{exa:NoComplaints}.
The threshold in such policies should also be determined in advance and not be dynamically changing, as it is difficult to explain why some orders got accepted while others got rejected.
Due to this reason, they do not consider the algorithm in \citet{han2015randomized} as practical in their setting.
Moreover, as we will see in Section~\ref{sec::computational}, even though the 1/2-competitive algorithm in \citet{han2015randomized} is theoretically the best-possible, its actual performance is often merely more than 1/2.
The practical performance of other algorithms seem to be better.

The data we observe is the sizes of all orders accepted by a greedy First-Come-First-Serve (FCFS) policy.
The sum of all observed order sizes for each $(\text{SKU}, \text{warehouse})$-pair is then at most the starting inventory, since the order sizes that cannot be fulfilled have been censored.
To create non-trivial instances, we re-scale the starting inventory amounts (which we know) for each SKU at each warehouse by a factor $\alpha\in[0,1]$, and test the performance of different accept/reject policies over different scaling factors $\alpha$.

There are multiple ways to implement our random-threshold policy.
The first implementation involves independently generating a random threshold for each SKU and for each warehouse.
The second implementation involves generating evenly-spaced percentiles of the threshold distribution $F$ across different warehouses.
More specifically, for each SKU and among all the $w$ warehouses, we take the $w$ thresholds defined by $F^{-1}(0)$, $F^{-1}(\frac{1}{w})$, ..., $F^{-1}(\frac{w-1}{w})$.
Then we randomly permute them and assign them over the $w$ warehouses, making accept/reject decisions at each warehouse based on the assigned threshold (scaled by the starting inventory).
The third implementation involves independently generating one single percentile for each SKU, and use this percentile for all warehouses.
These three implementations all have very similar performance.
We believe the second implementation, that is, the assignment of evenly-spaced percentiles to warehouses, is how our threshold distribution $F$ would be implemented in practice.
We then average the fulfillment ratios over the warehouses to determine the performance for a specific SKU.
We take an outer average over many independent random permutations of warehouses to define a final performance ratio for each of the 974 SKUs.

We find that the greedy FCFS policy has the best average-case performance ratio, even when the scaling factor $\alpha$ for initial inventory is small.
All three implementations of our random-threshold policies have similar performance.
The algorithm in \citet{han2015randomized} has the worst average-case performance ratio. 
While this is discouraging, we believe that the way in which order sizes are censored in our data favors FCFS, since large orders cannot come at the end.
Nonetheless, if we look at the worst case SKU, then our random-threshold policy has the best performance for a large fraction of scaling factors $\alpha$.
Indeed, our random-threshold policy being robust to the worst case is consistent with it having the best competitive ratio.
This robustness is not achieved by FCFS.

\subsubsection*{Roadmap.}
In Section~\ref{sec:Single} we introduce our results on a single knapsack.
We introduce the notations and then provide an overview of our analytical techniques in Section~\ref{sec:Techniques}. 
Section~\ref{sec:3/7alg} introduces the $\frac{3}{7} = 0.4285$-competitive algorithm relative to the optimal fractional packing, and Section~\ref{sec:0.4324alg} introduces the $0.4324$-competitive algorithm relative to the optimal integer packing.
Then in Section~\ref{sec::AdWords} we introduce our results on multiple knapsacks.
Section~\ref{sec:multipleALG} introduces the $0.2142$-competitive algorithm.
Section~\ref{sec:multipleUB} introduces an upper bound that no algorithm can be more than $0.4605$-competitive, which shows the fundamental challenge that knapsack constraint brings to the wholesale supply chain ordering problem.
Finally in Section~\ref{sec::computational}, we conduct computational study using real data from a Latin American chain department Store, and show the efficacy of threshold algorithms.

\section{Single Knapsack} 
\label{sec:Single}

In this paper we denote $[T] = \{1,2,...,T\}$ for any positive integer $T$.
We start with a single knapsack and let the capacity of the knapsack be $1$.
Let the entire set of items be indexed by $t \in [T]$, the sequence of its arrival.
For any $t \in [T], s_t$ refers to the size of item $t$.
The entire sequence of item sizes is then $S = (s_1, s_2, ..., s_T)$.
For any $A \subseteq [T]$, a subset of indices, let $\size(A) = \sum_{t \in A} s_t$ be the total size of items in $A$.

Suppose there is a clairvoyant decision maker who knows the entire sequence in advance.
This decision maker is going to take the \textit{optimal} actions (accept / reject) over the process.
Let this policy be $\OPT$.
Note that $\OPT$ does not necessarily guarantee to fill all the capacity of the knapsack, but it must be upper bounded by $1$.

For any specific sequence of $S$, let $\ALG(S)$ denote the total amount filled by $\ALG$ on this instance \textit{in expectation}, where expectation is taken over the randomness of the algorithm.
Here $\ALG$ is any generic algorithm, where in the following sections we will specify which algorithm it is by using slightly different notations for each algorithm.
Let $\OPT(S)$ denote the total amount filled by $\OPT$ on this sequence.
Note that 
\begin{align}
\OPT(S)=\max_{A\subseteq[T]: \size(A) \leq 1}\size(A).
\end{align}
We will also refer to a stronger optimum $\OPTp$ which is not only clairvoyant, but allowed to truncate items at will, with
\begin{equation}
\label{eqn:OPTp}
\OPTp(S)=\min\{s_1+\cdots+s_T,1\}
\end{equation}
It is self-evident that $\OPT(S)\le\OPTp(S)$ for any $S$.
We also use $\ALG, \OPT$ and $\OPTp$ for $\ALG(S), \OPT(S)$, and $\OPTp(S)$, respectively, if the sequence $S$ is clear from the context.

Under any policy, we say that an item $s_t$ is \textit{rejected} if it does not exceed the threshold $\tau$ of a threshold policy. If an item is rejected under a policy, we say that this policy rejects this item.
Under any policy, we say that an item $s_t$ is \textit{blocked} at the moment it arrives, if the remaining capacity of the knapsack is not enough for $s_t$ to fit in. If an item is blocked under a policy, we say that this policy blocks this item.
An item can be both rejected and blocked at the same time.

The focus of this paper is on randomized non-adaptive threshold algorithms.
We first define threshold algorithms as follows.
Let $\THR(\tau), \forall \tau \in [0,1]$ be a threshold algorithm that accepts any item whose size is greater or equal to $\tau$, as long as it can fit into the knapsack.
A $\THR(0)$ policy is also referred to as a greedy policy, $\Gre$: accept any item regardless of its size, as long as it can fit into the knapsack.
We will interchangeably use $\THR(0)$ and $\Gre$ for the same policy.

We say that a threshold algorithm is non-adaptive, if the threshold $\tau$ is determined before any item arrives, and does not depend on the future arriving items nor the past items observed.
Note that a threshold algorithm can be randomized, in which case $\tau$ is chosen from a probability distribution at the start and then fixed over time.

\subsection{Techniques for Analyzing Threshold Policies}
\label{sec:Techniques}

It is easy to see that randomization is necessary to achieve any non-zero competitive ratio.
Any deterministic algorithm, when faced with an initial item of a small size $\eps > 0$, must either accept or reject.
If it accepts, then it achieves a poor ratio when the item is followed by an item of size 1, packing only size $\eps$ whereas the optimal offline packing packs size 1.
On the other hand, if it rejects, then it achieves a poor ratio when the sequence ends after the first item, packing only 0 when the optimal offline packing packs size $\eps$.


\begin{figure}[!tb]\centering
\includegraphics[width=0.75\textwidth]{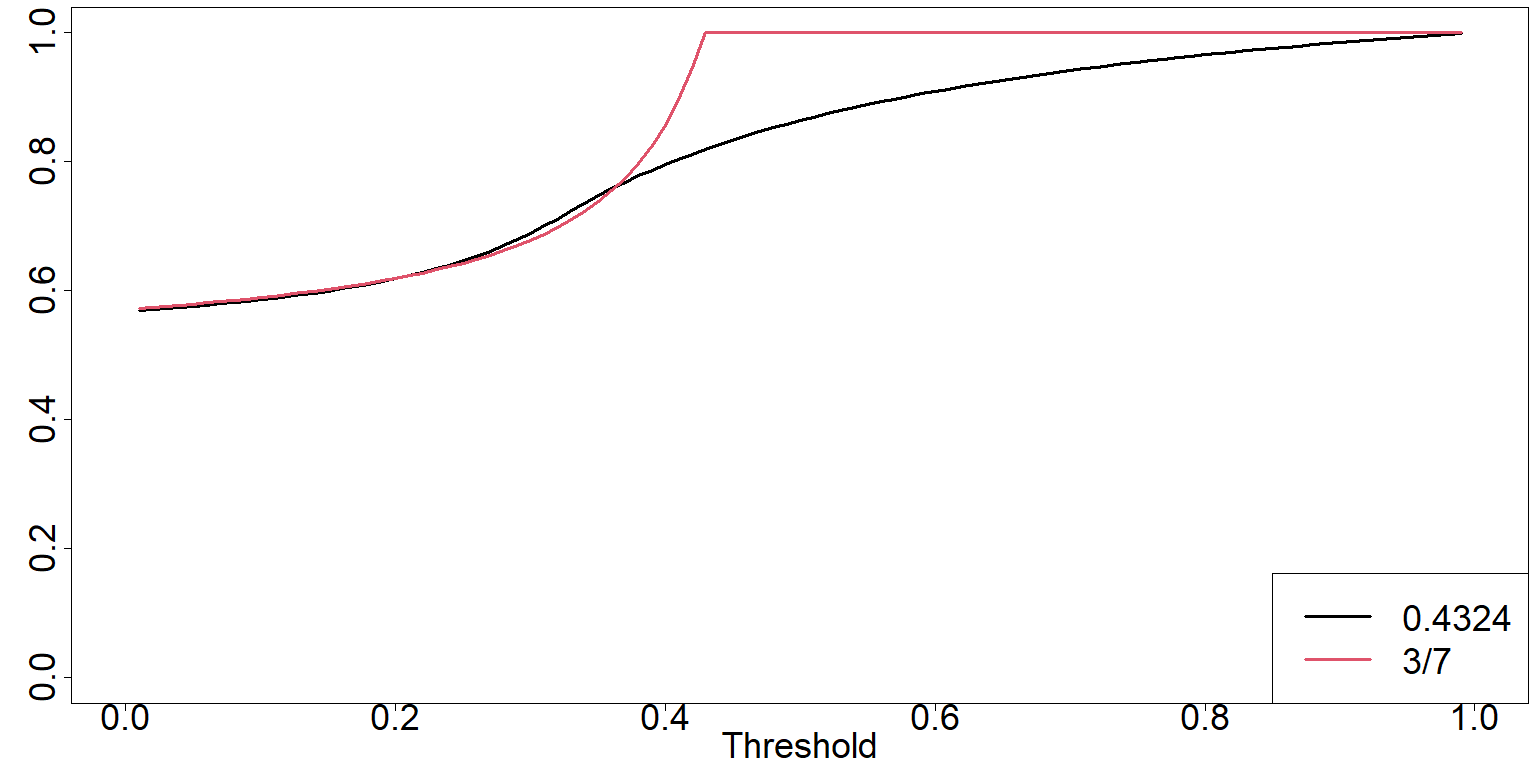}
\caption{Cumulative distribution functions of the thresholds from two random threshold algorithms}
\label{fig:CDFs}
\end{figure}

To achieve a non-zero competitive ratio, we need to design a random distribution of thresholds, which can be characterized by a Cumulative Distribution Function (CDF).
Figure~\ref{fig:CDFs} shows the two CDFs constructed for the $3/7\approx0.4285$-competitive algorithm relative to the optimal fractional packing, and for the $0.4324$-competitive algorithm relative to the optimal integer packing, respectively.
Below we provide intuitions how to optimize CDFs, and overview our analytical techniques.

We first start with the following simple randomized threshold algorithm to illustrate our idea of choosing the random distribution of thresholds.
This algorithm flips an initial biased coin.
With probability $2/3$, the algorithm sets $\tau = 0$ and greedily accepts any item which fits in the knapsack.
With probability $1/3$, the algorithm sets $\tau = 1/2$ and only accepts items whose sizes are at least $1/2$ (if such items exist).

We claim that this simple algorithm yields a constant competitiveness guarantee of 1/3.
To see why, first note that if the greedy policy can fit all the items, then it is optimal, and since the algorithm is greedy with probability 2/3, it would be at least 2/3-competitive.
Therefore, suppose that the greedy policy cannot fit some items, and consider two cases.
If the sequence contains no items of size at least 1/2, then the greedy policy must have packed size greater than 1/2 by the time it could not fit an item, and hence the algorithm packs expected size at least $2/3 \times 1/2 = 1/3$.
In the other case, let $m$ denote the size of the \textit{first} item to have size at least 1/2.
When the algorithm is not greedy, it packs size $m$; and when it is greedy, it packs size at least $\min\{m,1-m\}$, which equals $1-m$ because $m \geq 1/2$.  In expectation, the algorithm packs size at least
\begin{align*}
\frac{1}{3}m+\frac{2}{3}(1-m)=\frac{2}{3}-\frac{1}{3}m\geq\frac{1}{3}.
\end{align*}
Since the algorithm in both cases packs size at least 1/3, and the optimal offline packing cannot exceed 1, this completes the claim that the algorithm is $1/3$-competitive.

\begin{figure}[!tb]
    \begin{subfigure}{0.49\textwidth}
    \includegraphics[width=\textwidth]{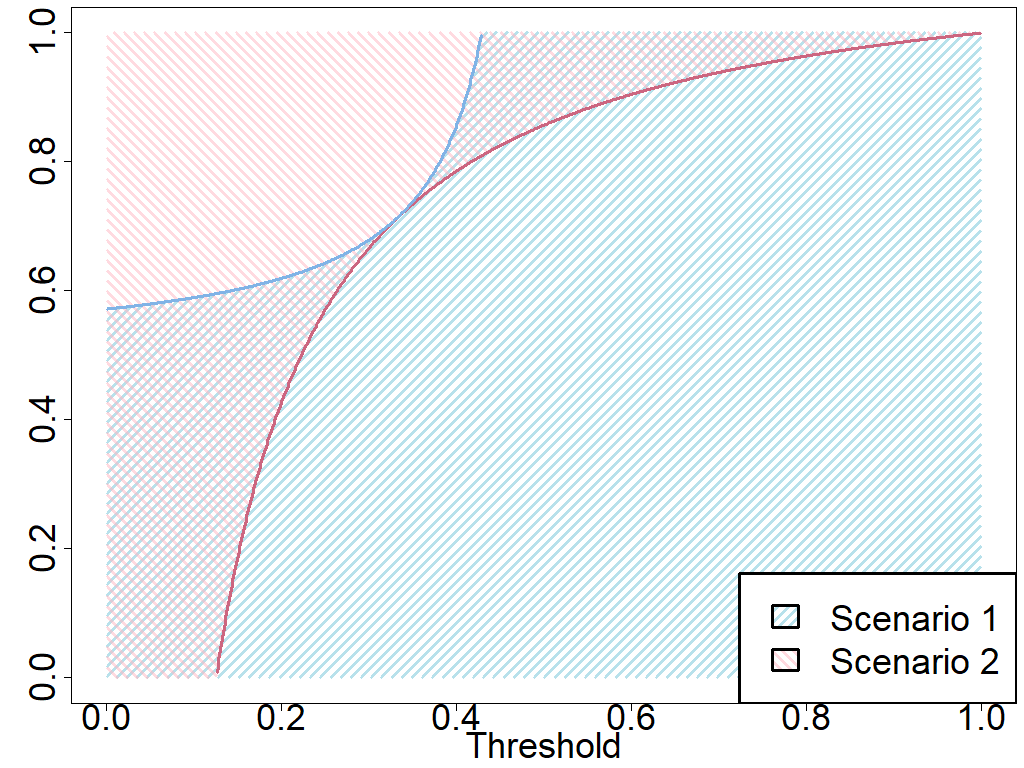}
    \caption{$0.4285$-competitive algorithm}
    \label{fig:Tangent37}
    \end{subfigure}
    \hfill
    \begin{subfigure}{0.49\textwidth}
    \includegraphics[width=\textwidth]{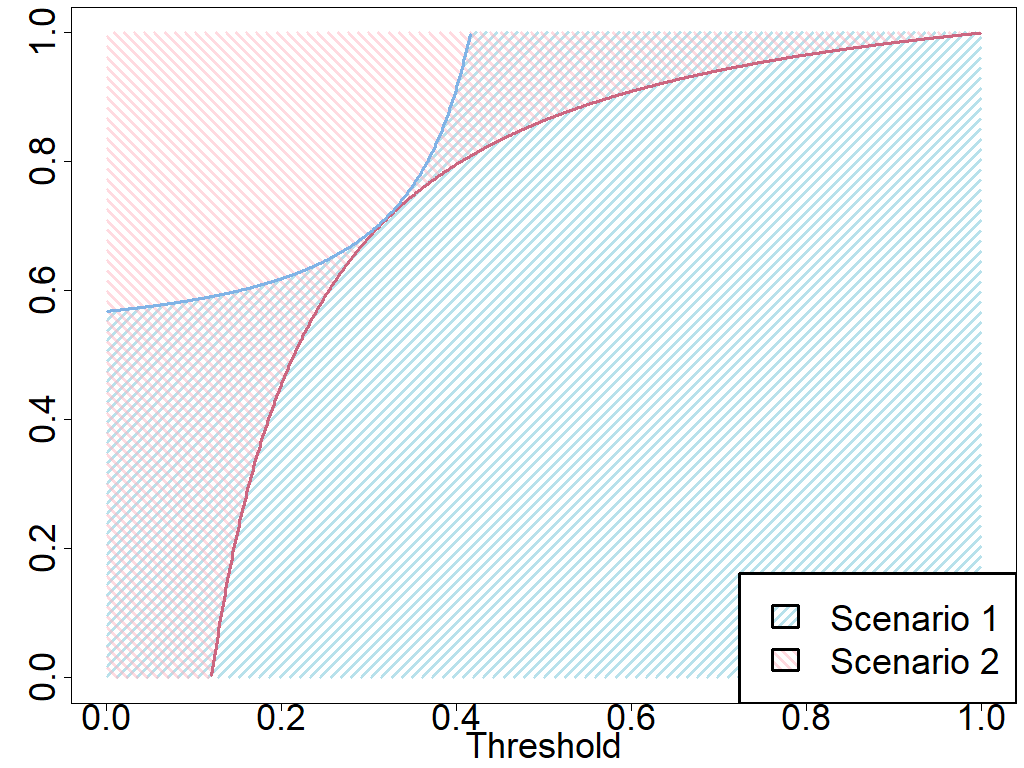}
    \caption{$0.4324$-competitive algorithm}
    \label{fig:Tangent}
    \end{subfigure}
    \hfill
{\small
\textit{Note:} The intersection is illustrated using both shading lines. Through comparing \ref{fig:Tangent37} and \ref{fig:Tangent}, the two algorithms have similar critical frontiers as illustrated using the solid blue and solid red curves.}
\caption{Cumulative distribution functions in two binding scenarios. }
\end{figure}

The previous algorithm effectively sets a random threshold where $\tau$ is a discrete probability distribution, that is, $\tau$ is equal to $0$ with probability $2/3$, and $1/2$ with probability $1/3$.
Now we generalize the above analysis to distributions characterized by CDF using parameter $c$, that is, $F(c,x)=\Pr(\tau \leq x)$, and optimize over the function $F(c,x)$.
The high-level intuition from the previous algorithm still carries to the new analysis, that there are two binding scenarios.
Scenario~1 is when accepting a smaller item earlier will block a larger item that arrives later, in which case we will need to set a larger threshold.
Scenario~2 is when there are only smaller items that arrive, in which case strategically waiting for larger items will end up accepting nothing.
We will need to set a smaller threshold to accept the smaller items.
For both scenarios, we can identify a family of worst case instances in that scenario, and evaluate the performance of any CDF on the respective worst case instance.

Below we illustrate this intuition on our $3/7\approx0.4285$-competitive algorithm when the fractional packing is the offline optimum.
Let $c \in [0,1/2)$ be any constant.
Scenarios~1 and~2 correspond to the blue and red areas in Figure~\ref{fig:Tangent37}.
In Scenario~1, there is a family of worst case instances (see expression \eqref{eqn::large_m} in the sketch proof of Theorem~\ref{thm:3/7}).
The dark blue curve illustrates a critical frontier for Scenario~1, defined as
\begin{align*}
F_1(c,x) = \frac{(1-c-x)}{1-2x},
\end{align*}
such that all the CDFs below this critical frontier will achieve at least $c$-competitiveness on this family of worst case instances.
Similarly, in Scenario~2, there is also a family of worst case instances explicitly given by $(\varepsilon, \varepsilon, ..., \varepsilon, x)$, where there are $\frac{1-x}{\varepsilon}+1$ many $\varepsilon$ at the beginning of this instance.
The dark red curve illustrates a critical frontier for Scenario~2, defined as
\begin{align}
F_2(c,x) = 2(1-c) - \frac{1-2c}{x}, \label{eqn:IllustratorScenario2}
\end{align}
such that all the CDFs above this critical frontier will achieve at least $c$-competitiveness on this family of worst case instances.
Any CDF that lives at the intersection of the two areas will jointly achieve at least $c$-competitiveness in both Scenarios.

To achieve a higher competitive ratio, we need to increase the value of $c$ in the critical frontier function $F_1(c,x)$, pushing the critical frontier for Scenario~1 towards the bottom-right corner. 
Intuitively, this means we are setting larger thresholds.
On the other hand, we need to increase the value of $c$ in the critical frontier function $F_2(c,x)$, pushing the critical frontier for Scenario~2 towards the top-left corner.
Intuitively, this means we are setting smaller thresholds.
The best performance is achieved when the two critical frontiers are tangent to each other. 
The parameter $c$ when two critical frontiers are tangent defines the best competitive ratio.

We further generalize this intuition to our $0.4324$-competitive algorithm when the integer packing is the offline optimum.
Let $c \in [3/7,1/2)$ be any constant.
Scenarios~1 and~2 correspond to the blue and red areas in Figure~\ref{fig:Tangent}.
The dark blue curve illustrates a critical frontier for Scenario~1, defined as
\begin{align}
F_3(c,x) = (1-c) - \frac{(1-2c)\ln{(1-x)}}{1-2x}, \label{eqn:IllustratorScenario1}
\end{align} 
such that all the CDFs below this critical frontier will achieve at least $c$-competitiveness on the worst case instance for Scenario~1.
Similarly, the dark red curve illustrates a critical frontier for Scenario~2.
The critical frontier for Scenario~2 under the $0.4324$-competitive algorithm is $F_2(c,x) = 2(1-c) - \frac{1-2c}{x}$, which is the same as that defined in \eqref{eqn:IllustratorScenario2} under the $0.4285$-competitive algorithm.
All the CDFs above this critical frontier will achieve at least $c$-competitiveness on the worst case instance for Scenario~2.
Any CDF that lives at the intersection of the two areas will jointly achieve at least $c$-competitiveness in both Scenarios.

In Figures~\ref{fig:Tangent37} and~\ref{fig:Tangent}, we see that both Scenarios~1 and~2 in the two figures are very similar.
However, the specific CDFs that we choose in the two settings are different.
This is because in Figures~\ref{fig:Tangent37} and~\ref{fig:Tangent} there are other critical frontiers that we do not illustrate.
Such unillustrated critical frontiers may live at the intersection of the two areas, and may or may not be tangent to the two illustrated critical frontiers.
Different CDFs that live at the intersection will have different performance; we compare them in Section~\ref{sec:AdditionalSimu} in the Online Appendix.

It is worth mentioning that, unlike traditional works in competitive analysis where the competitive ratio is found by optimizing a few parameters, we optimize over the cumulative distribution functions to find the competitive ratio.
We essentially solve an infinite dimensional optimization problem.
Solving infinite dimensional optimization problems in competitive and approximation ratio analysis has been used in the literature \citep{anunrojwong2022robustness, correa2021prophet, devanur2012online, huang2019online}.
Such problems are generally harder than optimizing over a few (finite-dimensional) parameters.

\subsection{A 0.4285-Competitive Algorithm Relative to the Optimal Fractional Packing} \label{sec:3/7alg}
We propose a randomized threshold policy, $\ALGN$, and prove it is $\frac{3}{7}$-competitive.
\begin{definition}
\label{defn:3/7alg}
Let $\ALGN$ be a randomized threshold policy that runs as follows,
\begin{enumerate}
\item At the beginning of the entire process, randomly draw $\tau$ from a distribution whose cumulative distribution function (CDF) is given by
\begin{equation}
\label{eqn:defineF}
\FN(x) = \left\{
\begin{aligned}
& \frac{4/7-x}{1-2x}, & \quad x \in [0, 3/7]\\
& 1, & \quad x \in (3/7,1]
\end{aligned}
\right.
\end{equation}
\item We apply $\THR(\tau)$ policy throughout the process.
\end{enumerate}
\end{definition}
Notice that $\FN(0) = 4/7$.
This is the point mass we put on $\tau=0$.
This means that with probability $4/7$, we will perform $\Gre$.

Since $\tau$ is chosen at the start of the horizon and fixed throughout, $\ALGN$ is a non-adaptive threshold policy.
Note that our desired algorithm does not know how many items are there in total, nor does it know the sizes of the items.

Now we state and prove our first result.
\begin{theorem}
\label{thm:3/7}
$$\inf_{S} \frac{\ALGN(S)}{\OPTp(S)} \geq \frac{3}{7}$$
\end{theorem}

We sketch the proof idea below, and defer the details to Section~\ref{sec:proof:thm:3/7} in the Online Appendix.

\proof{Sketch Proof of Theorem~\ref{thm:3/7}.}
Note that the CDF that we constructed in Definition~\ref{defn:3/7alg} puts a point mass on 0, meaning that with $F(0)$ probability the algorithm mimics $\Gre$.
Let $m$ denote the size of the smallest item which $\Gre$ does not fit.
For example, in an arrival sequence $S_0 = (0.05, 0.2, 0.15, 0.5, 0.6, 0.4)$, the items that $\Gre$ does not fit are $\{0.6, 0.4\}$, and the smallest item that greedy does not fit is of size $m=0.4$.

In the case where $m<1/2$, the algorithm packs expected size at least
\begin{align} \label{eqn::small_m}
F(0)(1-m)+(F(m)-F(0))m.
\end{align}
This expression contains two components.
The first component suggests that with probability $F(0)$, at least $(1-m)$ amount is packed into the knapsack.
The second component suggests that with probability $(F(m) - F(0))$, $m$ is packed into the knapsack.

In the case where $m\ge1/2$, the analysis is more challenging.
We define $q$ to be the maximum size such that, at the time of arrival of the item of size $m$, it would not fit even if we could ``magically discard'' every accepted item of size (strictly) less than $q$.
In the same example when the arrival sequence is $S_0$, if we discard items $0.05$ and $0.15$ then $m=0.4$ still does not fit; but if we further discard $0.2$ then $m=0.4$ would fit.

By the maximality of $q$, there must exist an item of size $q$.
After carefully analyzing the cases (including the one where $q>m$), we show that the algorithm's expected packing size is minimized in the case where it equals $q$ when the threshold is at most $q$, and $1-q+\eps$ (for an arbitrarily small $\eps>0$) when the threshold is greater than $q$.  Therefore, it is lower-bounded by
\begin{align} \label{eqn::large_m}
F(q)q+(1-F(q))(1-q).
\end{align}
Finally, we solve for the maximum $c$ at which there exists a threshold distribution $F$ such that both expressions~\eqref{eqn::small_m} and~\eqref{eqn::large_m} exceed $c$ (for all $m$ and $q$).
This turns out to be $c=3/7\approx0.4285$, and since the optimal \textit{fractional} packing cannot exceed 1, the corresponding random-threshold algorithm is 0.4285-competitive relative to the stronger, fractional optimum.
\Halmos
\endproof

\subsubsection{Tightness proof of the $0.4285$-competitive algorithm.} \label{sec::tight37}

In this section we show that the guarantee of $\inf_S\frac{\ALGN(S)}{\OPTp(S)}\ge\frac{3}{7}$ from Definition~\ref{defn:3/7alg} is best-possible, relative to $\OPT^+$, among all randomized threshold policies.
To do this, we invoke the minimax theorem of \citet{yao1977probabilistic}, which says that it suffices to construct a distribution $\cS$ over sequences $S$ for which
\begin{align*}
\sup_{\ALG:\ALG=\THR(\tau),\tau\in[0,1]}\frac{\bE_{S\sim\cS}[\ALG(S)]}{\bE_{S\sim\cS}[\OPTp(S)]}\le\frac{3}{7}.
\end{align*}
In particular, we only need to establish that $\frac{\bE_{S\sim\cS}[\ALG(S)]}{\bE_{S\sim\cS}[\OPTp(S)]}\le\frac{3}{7}$ for deterministic threshold policies specified by a $\tau\in[0,1]$.


\begin{theorem}
\label{thm:3/7example}
There exists a distribution $\cS$ over arrival sequences $S$ such that for any $\tau\in[0,1]$, the algorithm $\ALG=\THR(\tau)$ has
$
\frac{\bE_{S\sim\cS}[\ALG(S)]}{\bE_{S\sim\cS}[\OPTp(S)]}\le\frac{3}{7}.
$
\end{theorem}

\proof{Proof of Theorem~\ref{thm:3/7example}.}
Let the random arrival sequence be $S$:
\begin{equation}
S = \left\{
\begin{aligned}
& (1/3, 2/3+\eps), && \text{with prob. } 3/7; \\
& (\underbrace{\eps, \eps, ..., \eps}_{2/ (3\eps) + 1 \text{ many}}, 1/3), && \text{with prob. } 3/7; \\
& (\eps, 1), && \text{with prob. } 1/7;
\end{aligned}
\right. \label{eqn:3/7example}
\end{equation}

Following each realization of $S$, $\OPTp(S)=1$. So we have $\bE_S[\OPTp(S)] = 1$.

For any $\ALG=\THR(\tau),\tau\in[0,1]$, we enumerate all the potential values of $\tau$ in the following.\\
\textbf{Case 1}: $0\le\tau\le\eps$. In this case,
\begin{align*}
\bE_S[\ALG(S)] = \frac{1}{3} \cdot \frac{3}{7} + \left(\frac{2}{3}+\eps\right) \cdot \frac{3}{7} + \eps \cdot \frac{1}{7} = \frac{3}{7} + \frac{4}{7} \cdot \eps
\end{align*}
\textbf{Case 2}: $\eps < \tau \leq 1/3$. In this case,
\begin{align*}
\bE_S[\ALG(S)] = \frac{1}{3} \cdot \frac{3}{7} + \frac{1}{3} \cdot \frac{3}{7} + 1 \cdot \frac{1}{7} = \frac{3}{7}
\end{align*}
\textbf{Case 3}: $1/3 < \tau \leq 2/3+\eps$. In this case,
\begin{align*}
\bE_S[\ALG(S)] = \left(\frac{2}{3}+\eps\right) \cdot \frac{3}{7} + 0 \cdot \frac{3}{7} + 1 \cdot \frac{1}{7} = \frac{3}{7} + \frac{3}{7} \cdot \eps
\end{align*}
\textbf{Case 4}: $2/3+\eps < \tau \leq 1$. In this case,
\begin{align*}
\bE_S[\ALG(S)] = 1 \cdot \frac{1}{7} = \frac{1}{7}
\end{align*}

In all, we have enumerated all the values that a threshold can take.
In all cases, the performance of the threshold $\THR(\tau)$ policy has an expected performance of no more than $3/7+4/7 \cdot \eps$.
But $\bE_S[\OPTp(S)] = 1$.
By taking $\eps \to 0^+$ we finish the proof.
\Halmos\endproof

\subsection{A 0.4324-Competitive Algorithm Relative to the Optimal Integer Packing} \label{sec:0.4324alg}

In this section we are going to introduce a threshold policy that achieves the best-possible competitive ratio in the non-adaptive threshold family. In Section~\ref{sec:0.4324alg:example} we will show it is best-possible.

We first define some parameters that are going to be useful in the following analysis.
Let $H: [3/7,1/2) \times (0,1/2) \to \mathrm{R}_+$ be a bivariate real function defined as follows,
\begin{align*}
H(c,x) = & \frac{1-2c}{x} - \frac{(1-2c)\ln{(1-x)}}{1-2x} - (1-c) \\
= & (1-2c) \cdot \left( \frac{1}{x} - \frac{\ln{(1-x)}}{1-2x} \right) - (1-c).
\end{align*}
Intuitively, $H(c,x) = F_3(c,x) - F_2(c,x)$ is the difference between the two critical frontier functions that we have defined in \eqref{eqn:IllustratorScenario1} and \eqref{eqn:IllustratorScenario2} in Section~\ref{sec:Techniques}.

\begin{figure}[!tb]
\centering
\includegraphics[width=0.75\linewidth]{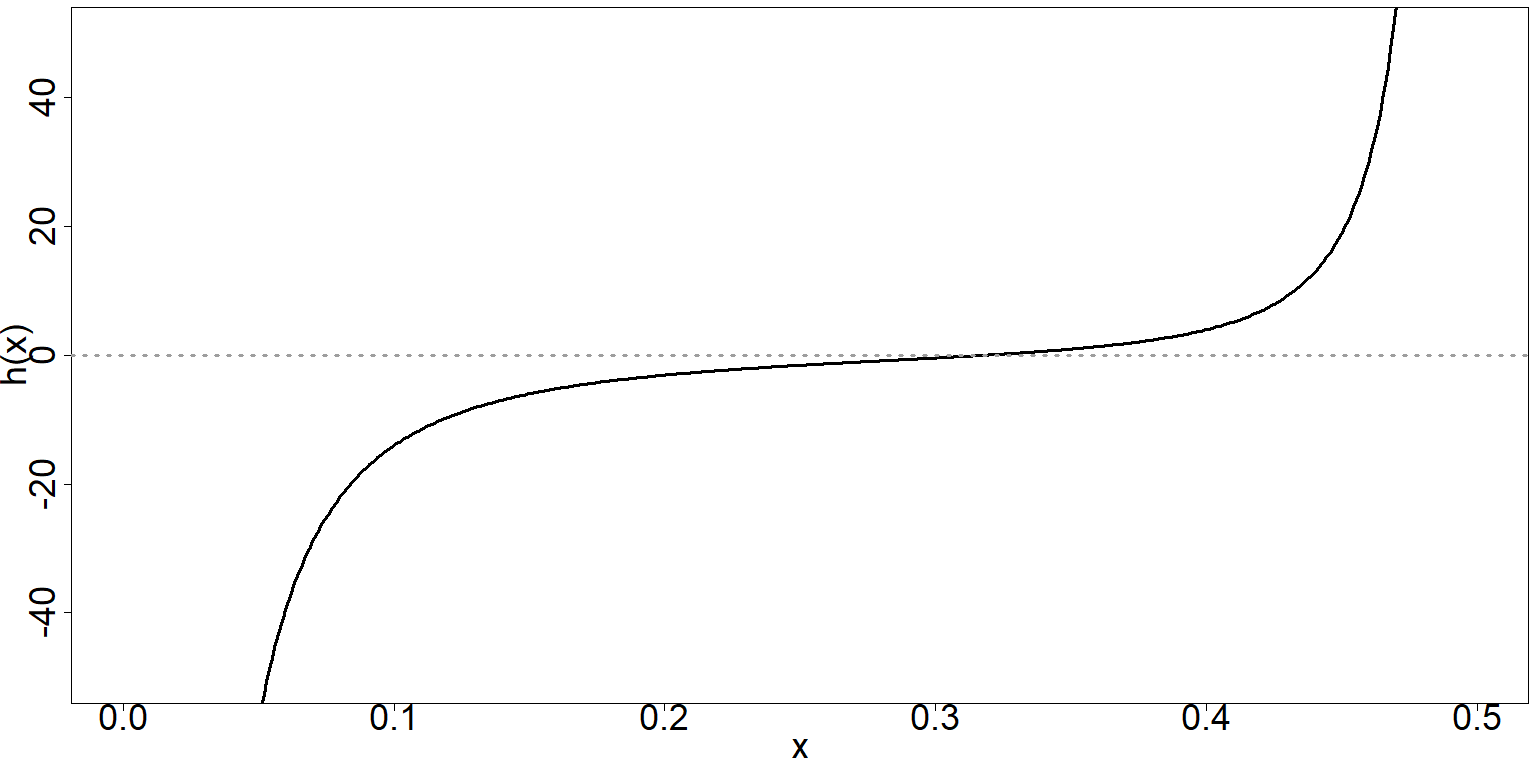} 
\caption{ An illustration of $\partial H(c,x) / \partial x$, plotted between $(0,1/2)$. }
\label{fig:dHcx_dx}
\end{figure}

Now fix $c$ to be any number in $[3/7,1/2)$. 
Define $\qG$ to be the unique local minimizer on the second coordinate of $H(c,x)$ between $(0,1/2)$.
That is, let $\qG$ be the only solution to the following function,
\begin{equation*}
h(x) = \frac{\partial H(c,x)}{\partial x} = \frac{(1-2c)\cdot\Big(2x^3-7x^2+5x-1-2(1-x)x^2\ln{(1-x)}\Big)}{x^2(1-x)(1-2x)^2} = 0,
\end{equation*}
or, approximately, $$\qG \approx 0.3185.$$
See Figure~\ref{fig:dHcx_dx} for an illustration of the unique solution.

Define $\cG$ to be the only solution in $[3/7,1/2)$, such that
\begin{equation}
\label{eqn:H=0}
H(\cG,\qG) = \frac{1-2\cG}{\qG} - \frac{(1-2\cG)\ln{(1-\qG)}}{1-2\qG} - (1-\cG) = 0,
\end{equation}
or, approximately, $$\cG \approx 0.4324.$$

We can check the following inequality:
$\forall q \in (0,1/2)$,
\begin{equation}
\label{eqn:defineHproperty1}
H(\cG,q) \geq H(\cG, \qG) = 0
\end{equation}


We propose another randomized threshold policy, $\ALGG$, using another random threshold. 
It gives us an improved $0.4324$-competitive guarantee.
\begin{definition}
\label{defn:0.4324alg}
Let $\ALGG$ be a randomized threshold policy that runs as follows,
\begin{enumerate}
\item At the beginning of the entire process, randomly draw $\tau$ from a distribution whose CDF is given by
\begin{equation}
\label{eqn:defineF2}
\FG(x) = \left\{
\begin{aligned}
& (1-\cG) - \frac{(1-2\cG) \ln{(1-x)}}{1-2x}, & x \in [0, \qG]\\
& 2(1-\cG) - \frac{1-2\cG}{x}, & x \in (\qG,1]
\end{aligned}
\right.
\end{equation}
\item We apply $\THR(\tau)$ policy throughout the process.
\end{enumerate}
\end{definition}
Notice that $\FG(0) = 1-\cG$.
This is the point mass we put on $\tau=0$.
This means that with probability $1-\cG \approx 0.5676$, we will perform $\Gre$.

We state our main result here.
\begin{theorem}
\label{thm:0.4324}
$$\inf_{S} \frac{\ALGG(S)}{\OPT(S)} \geq \cG \approx 0.4324$$
\end{theorem}

The proof idea is the same as in Theorem~\ref{thm:3/7}, but in order to improve it, we are more careful in upper bounding the performance of $\OPT$. 
The way to bound $\OPT$ is by noting that even the optimal algorithm cannot guarantee to always pack the knapsack fully.
To compare to the proof of Theorem~\ref{thm:3/7}, Case 1.2 will be different.
The proof details are deferred to Section~\ref{sec:0.4324alg:proof} in the Online Appendix.

\subsubsection{Tightness proof of the $0.4324$-competitive algorithm.} \label{sec:0.4324alg:example}

In this section we show that the guarantee of $\inf_S\frac{\ALG(S)}{\OPT(S)}\ge \cG$ from Definition~\ref{defn:0.4324alg} is best-possible among all randomized threshold policies.
As in Section~\ref{sec::tight37}, we invoke the minimax theorem of \citet{yao1977probabilistic}, which says that it suffices to construct a distribution $\cS$ over sequences $S$ for which
\begin{align*}
\sup_{\ALG:\ALG=\THR(\tau),\tau\in[0,1]}\frac{\bE_{S\sim\cS}[\ALG(S)]}{\bE_{S\sim\cS}[\OPT(S)]}\le \cG.
\end{align*}


\begin{theorem}
\label{thm:0.4324example}
There exists a distribution $\cS$ over arrival sequences $S$ such that for any $\tau\in[0,1]$, the algorithm $\ALG=\THR(\tau)$ has
$
\frac{\bE_{S\sim\cS}[\ALG(S)]}{\bE_{S\sim\cS}[\OPT(S)]}\leq \cG \approx 0.4324.
$
\end{theorem}

We prove Theorem~\ref{thm:0.4324example} by carefully selecting a distribution of worst case arrival sequences.
This distribution is similar to the one defined in \eqref{eqn:3/7example}.
But instead of being a discrete probability distribution, it involves a continuum of instances that conform to a continuous probability distribution.
The detailed construction of the distribution and the proof of Theorem~\ref{thm:0.4324example} are deferred to Section~\ref{sec:thm:0.4324example:proof} in the Online Appendix.

\section{Multiple Knapsacks} \label{sec::AdWords}

In this section we generalize our results to multiple knapsacks.
We define the problem here, then in Section~\ref{sec:multipleALG} we introduce the $0.2142$-competitive algorithm, and in Section~\ref{sec:multipleUB} we introduce the impossibility result that no algorithm can yield more than $0.4605$-competitiveness.

We manage $N$ divisible knapsacks indexed as $j \in [N]$, each having size $B_{j}, \forall j\in[N]$.
In each period of time, one item $t \in [T]$ arrives with an associated vector of $N$ sizes $(s_{t1}, s_{t2},...,s_{tN}) \in \bR_{+}^N$.
The sizes are revealed upon arrival, and each item must immediately be either entirely accepted by one knapsack, in which case $s_{tj}$ amount is filled up in knapsack $j$, or entirely rejected (there is no partial fulfillment).
The objective is to maximize the sum of sizes of accepted items from all knapsacks, i.e. maximize the space in the knapsacks filled.
Below we explain how to capture Example~\ref{exa:Cocoa} using our multi-knapsack model.

\begin{example}[Cocoa distribution continued]
\label{exa:CocoaContinued}
Let $A_j$ denote the initial amount of specification $j$. 
Each specification $j$ has a fixed market price $p_j$ per unit. 
Request $t$ from downstream retailers arrive in sequence, which can be satisfied by a given amount $a_{tj}$ of each specification $j$. 
The manufacturer chooses at most one specification with sufficient amount remaining to satisfy the request, noting that mixing specifications is disallowed.
The objective is to maximize the revenue at the end of the horizon, which equals the sum across specifications of amount sold times unit price $p_j$.
To capture this using our model, we set $B_j = A_j p_j$, and $s_{tj} = a_{tj} p_j$.
\end{example}

We compare the algorithm's performance relative to the space filled by an optimal offline packing, who knows the entire sequence of items in advance.
This generalization can be seen as a modification of the AdWords budget allocation problem as in \citet{mehta2005adwords}, where we do not allow the partial allocation of any queries that go over budget.


\subsection{A 0.2142-Competitive Algorithm}\label{sec:multipleALG}

We first overview the AdWords problem originally proposed in \citet{mehta2005adwords}.
The language we use are from the book by \citet{mehta2013online}.
In period $t$, an item $t$ arrives with an associated vector of $N$ sizes $(s_{t1}, s_{t2},...,s_{tN}) \in \bR_{+}^N$.
Suppose that, at this moment, some $b_{tj}$ amount of space has been filled in each knapsack $j \in [N]$.
If we assign the item to knapsack $j$, then $\min\{B_j-b_{tj}, s_{tj}\}$ amount of stock from knapsack $j$ will be filled -- we allow for \textbf{truncation} in the AdWords problem.
For this AdWords problem, the following greedy algorithm is well-known.

\begin{definition}[Algorithm 8, \citet{mehta2013online}]
\label{defn:AdWordsGreedy}
When item $t$ arrives, find the knapsack such that $\tilde{j} \in \argmax_{j \in [N]} \min\{s_{tj}, B_j - b_{tj}\}$, and fit the item to knapsack $\tilde{j}$.
\end{definition}


It is well known that the greedy algorithm defined above achieves a competitive ratio of $1/2$.
For any instance $S$, let $\ALG_\mathsf{AW}(S)$ denote the total amount filled by the greedy algorithm from Definition~\ref{defn:AdWordsGreedy}, which is allowed to truncate.
Let $\OPT_\mathsf{AW}(S)$ denote the total amount filled by a clairvoyant decision maker, which is also allowed to truncate.
When $S$ is clear from the context, we drop $S$ from the notations.
\begin{proposition}[Theorem 5.1, \citet{mehta2013online}]
\label{prop:AdWordsGreedy}
The greedy algorithm from Definition~\ref{defn:AdWordsGreedy} is $1/2$-competitive for the AdWords problem, i.e. $\forall S$, $$\ALG_\mathsf{AW}(S) \geq \frac{1}{2} \OPT_\mathsf{AW}(S).$$
\end{proposition}

Now we return to our multiple knapsack problem without truncation.
We will leverage a definition of ``phantom'' capacity, which is similar in concept to the algorithmic ingredients in online matching with stochastic rewards, \citealp[see, e.g.,][Algorithm~2]{goyal2023online}, and \citealp[][Algorithm~1]{huang2020online}.

\begin{definition}
\label{defn:dependentSize}
We define our proposed algorithm, which essentially combines the two algorithms from Definitions~\ref{defn:3/7alg} and \ref{defn:AdWordsGreedy}. Let $d_{tj}$ denote the ``phantom'' capacity filled in each knapsack $j \in [N]$ at time $t\in[T]$, if all the items routed to knapsack $j$ were accepted, with truncation allowed. The algorithm proceeds as follows:
\begin{enumerate}
\item Initialization: at time $t=0$, set $b_{1j} = 0, d_{1j} = 0, \forall j \in [N]$; for each knapsack $j$, draw a random threshold $\tau_j, \forall j \in [N]$ using the CDF defined in Definition~\ref{defn:3/7alg}, where we properly re-scale the support to be $[0,B_j]$.
\item For each item $t$, find $\tilde{j} \in \argmax_{j \in [N]} \min\{s_{tj}, B_j - d_{tj}\}$, route item $t$ to knapsack $\tilde{j}$, and update (i) $d_{(t+1) \tilde{j}} = d_{t \tilde{j}} + s_{t \tilde{j}}$; (ii) $d_{(t+1) j} = d_{t j}, \forall j \ne \tilde{j}$.
\item If $\tau_{\tilde{j}} \leq s_{t \tilde{j}} \leq B_j - b_{t \tilde{j}}$, accept item $t$ by matching it to knapsack $\tilde{j}$, and update (i) $b_{(t+1) \tilde{j}} = b_{t \tilde{j}} + s_{t \tilde{j}}$; (ii) $b_{(t+1) j} = b_{t j}, \forall j \ne \tilde{j}$.
\end{enumerate}
\end{definition}

We distinguish between $b_{jt}$ the actual amount filled by our policy, which we only update once we accept an item, and $d_{jt}$ the phantom amount filled by the \citet{mehta2013online} routing policy, which we update once we route an item (but not necessarily resulting in acceptance).

We use the algorithm from Definition~\ref{defn:AdWordsGreedy} to route items to knapsacks, and then use the single-knapsack algorithm to decide if we actually accept it.
We do not prescribe the correlations between the thresholds for each knapsack.
They can be independent, and they can also be arbitrarily correlated.

For any instance $S$, let $\ALG(S)$ denote the total amount filled by the combined algorithm from Definition~\ref{defn:dependentSize}, which is not allowed to truncate.
Let $\OPT(S)$ denote the total amount filled by a clairvoyant decision maker, which is also not allowed to truncate.
Using these notations, we present Theorem~\ref{thm:dependentSize}, whose proof is deferred to Section~\ref{sec:app:dependentSize} in the Online Appendix.

\begin{theorem}
\label{thm:dependentSize}
The combined algorithm from Definition~\ref{defn:dependentSize} is $\frac{3}{14}$-competitive, i.e. $$\inf_{S} \frac{\ALG(S)}{\OPT(S)} \geq \frac{3}{14}.$$ 
%
\end{theorem}

The algorithmic idea of phantom capacity as in Definition~\ref{defn:dependentSize} enables us to combine the algorithms from Definitions~\ref{defn:3/7alg} and~\ref{defn:AdWordsGreedy}, and to combine the results from Theorem~\ref{thm:3/7} and Proposition~\ref{prop:AdWordsGreedy}.
In fact, the above algorithmic idea and proof framework allow us to combine \textbf{any} AdWords algorithm with \textbf{any} single-knapsack algorithm relative to the optimal fractional packing.
We generalize Theorem~\ref{defn:dependentSize} by stating the following result, whose formal statement and proof are deferred to Section~\ref{sec:app:dependentSize} in the Online Appendix.

\begin{theorem}
\label{thm:generalization}
Let $\ALG_{\mathsf{AW}}$ be any algorithm that is $\Gamma_{\mathsf{AW}}$-competitive for the AdWords problem;
let $\ALG_{\mathsf{SK}}$ be any algorithm that is $\Gamma_{\mathsf{SK}}$-competitive for the single knapsack problem relative to the optimal fractional packing.
If we combine $\ALG_{\mathsf{AW}}$ and $\ALG_{\mathsf{SK}}$, the combined algorithm is $(\Gamma_{\mathsf{AW}}\Gamma_{\mathsf{SK}})$-competitive.
\end{theorem}

Following Theorem~\ref{thm:generalization}, if we combine the $0.5016$-competitive algorithm in \citet{huang2020adwords} with the $0.5$-competitive algorithm in \citet{han2015randomized}, we straightforwardly obtain a $0.2508$-competitive algorithm.
Yet this combined policy, because it uses the algorithm in \citet{han2015randomized}, does not belong to the family of threshold algorithms for a single knapsack.
Moreover, Theorem~\ref{thm:generalization} allows us to use any advancement in the AdWords problem to improve the performance of our unit-density knapsack problem.

%


\subsection{An Upper Bound for Multiple Knapsacks Strictly Less Than 0.5} \label{sec:multipleUB}

Fix a small $\eps>0$.
There are $N$ knapsacks with sizes $B_1=\cdots=B_N=1$.
The arrival sequence $S$ deterministically starts with $N$ items each of which take size $\eps$ in a particular knapsack, and size 0 in all other knapsacks.
Specifically,
\begin{align*}
s_{t,t'}=\begin{cases}
\eps, &\text{if }t=t' \\
0, &\text{otherwise}
\end{cases}
&&\forall\ t,t'\in[N].
\end{align*}

After item $N$, with probability $\alpha$, the arrival sequence terminates; with probability $1-\alpha$, there are $N$ more items whose sizes adhere to a  ``upper-triangular graph'', defined as follows.
A permutation $\pi:[N]\to[N]$ is chosen uniformly at random among all $N!$ possibilities.
Item $N+1$ takes size 1 in all knapsacks.
Item $N+2$ takes size 1 in all knapsacks, except knapsack $\pi(1)$, where it takes size 0.
Item $N+3$ takes size 1 in all knapsacks, except knapsacks $\pi(1)$ and $\pi(2)$, where it takes size 0.
This construction is repeated until item $2N$, which takes size 1 only in knapsack $\pi(N)$.
Formally,
\begin{align*}
s_{N+t,t'}=\begin{cases}
1, &\text{if }t'\notin\{\pi(1),\ldots,\pi(t-1)\} \\
0, &\text{if }t'\in\{\pi(1),\ldots,\pi(t-1)\}
\end{cases}
&&\forall\ t,t'\in[N].
\end{align*}

The optimal solution matches items $1,\ldots,N$ to their corresponding knapsacks if the arrival sequence terminates after item $N$, and rejects items $1,\ldots,N$ otherwise, matching items $N+1,\ldots,2N$ to knapsacks $\pi(1),\ldots,\pi(N)$, respectively, instead.  Therefore
\begin{align} \label{eqn::upperTriangular}
\bE_S[\OPT(S)]=(\alpha)N\eps+(1-\alpha)N.
\end{align}

Meanwhile, the algorithm does not know whether the arrival sequence will terminate after item $N$, nor does it know $\pi$.
In the first phase, any algorithm can be captured by how many $\eps$'s it accepts, which we denote using $e \in \{0,1,...,N\}$.
Should the second phase occur, by the symmetry of the random permutation, an algorithm cannot do better than placing an arriving item arbitrarily into an empty knapsack (where it will take size 1) whenever possible.  Therefore,
the expected reward of any algorithm when the second phase does occur is completely determined by $e$.
By studying the different cases characterized by $e$, we can derive an upper bound on the competitive ratio.
The proof of Theorem~\ref{thm:knapsackDEexample} is deferred to Section~\ref{sec:app:knapsackDEexample} in the Online Appendix.


\begin{theorem}
\label{thm:knapsackDEexample}
There exists a distribution $\cS$ over arrival sequences $S$ such that for any (adaptive or non-adaptive) deterministic algorithm $\ALG$, we have
\begin{align*}
\frac{\bE_{S\sim\cS}[\ALG(S)]}{\bE_{S\sim\cS}[\OPT(S)]}\le\frac{35}{76} \approx 0.4605.
\end{align*}
By Yao's minimax theorem, the competitive ratio of any (possibly randomized) algorithm cannot be greater than 35/76.
\end{theorem}

\section{Simulations Using Data from a Latin American Department Store} \label{sec::computational}

We use supply chain data from a Latin American chain department store to computationally study the performance of our algorithms.
The supply chain data contains 974 SKUs, and their associated order quantities from different local stores to a total of 21 regional warehouses.

The category that we focus on is young women's fashion products.
Since fashion products are highly unpredictable in its sales, we adopt the lens of competitive analysis, which is natural when there is no knowledge about future arrival sequences.
There is typically only 1 selling season, and the selling season typically lasts for 3-6 months.
At the beginning of the selling season, there is an initial stock placed in each regional warehouse.
There is no inventory replenishment throughout the selling horizon.
Orders to a specific warehouse cannot be split or redirected to a different warehouse, because of the long shipment time and the high managerial costs involved.
Therefore, our industry partner faces the same accept/reject problem on item sizes, and has the same goal of maximizing total inventory fulfilled, equal to the sum of sizes of accepted orders.

To give a concrete example, here is an arrival sequence in the winter between year 2015 and 2016, for one SKU of women purse.
\begin{equation}
\label{eqn:example}
S=(7,18,80,41,1,30,12,17).
\end{equation}
This selling season spans the local black Friday, Christmas, and New Year.
And the item quantities are in commercial units.

Sequence $S$ is an observed sequence of item sizes, accepted by greedy (FCFS) in the real world supply chain.
The sum of all the item sizes is smaller than the starting inventory, which, in this case, is $208$ units.
Any item which could not have been fulfilled are censored from the data, and the true sequence of item sizes might have been larger.
Because of this potential censoring of larger sizes, we create non-trivial arrival instances by re-scaling the starting inventory amounts (which we know) for each SKU at each warehouse by a factor $\alpha\in[0,1]$, and then test the performance of different accept/reject policies over different scaling factors $\alpha$.
This is a limitation of our computational study.
Nonetheless, we believe that this censoring only favors FCFS in our computational study, because large orders rarely appear in the end of a sequence (which would cause FCFS to perform poorly).
Such an experiment setup to vary the initial inventory level is common in the online decision making literature; see \citet{zhang2005revenue, liu2008choice}.

\begin{figure}[!tb]
\centering
\includegraphics[width=0.95\linewidth]{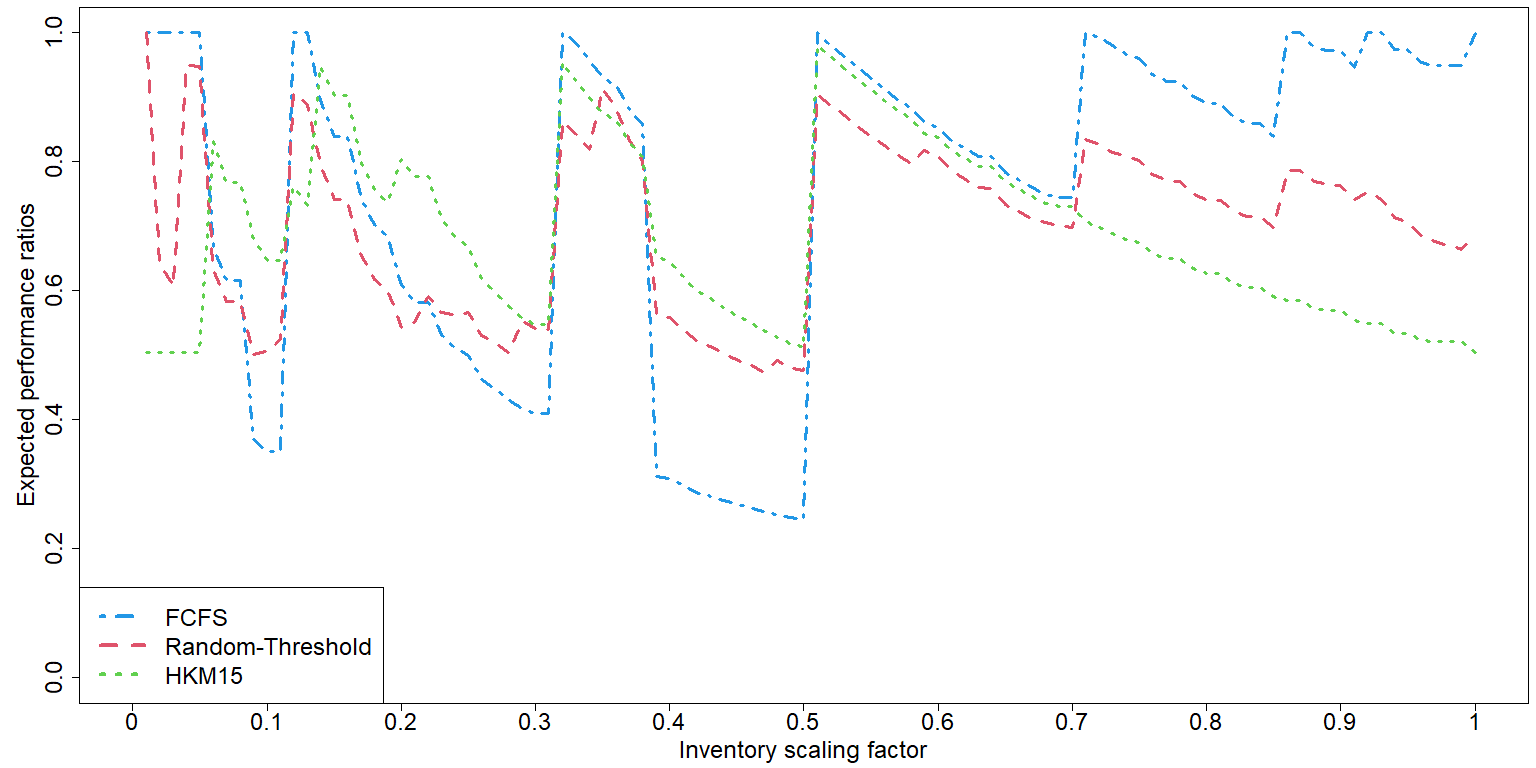} 
\caption{Computational performance on one real arrival sequence shown in Equation~\eqref{eqn:example}.}
\label{fig:computational}
\end{figure}

We compute the expected revenue from our proposed random threshold algorithm from Definition~\ref{defn:3/7alg}, named \textbf{Random-Threshold}; the (deterministic) revenue from first-come-first-serve policy, named \textbf{FCFS}; and the expected revenue from the algorithm suggested in \citet{han2015randomized}, named \textbf{HKM15}.
The results are shown in Figure~\ref{fig:computational}, where we have divided all the numbers by its corresponding offline optimal integer packing.
The offline optimal packing serves as an upper bound, so that the performance ratio is always between $0$ and $1$, with higher ratios indicating better performance.

When inventory level is either very small or very large, FCFS achieves near-optimal performance.
This is not surprising, because FCFS always tries to accept an order if possible: when inventory is large then FCFS could almost accept everything except for the ones that arrive late.
On our specific arrival instance \eqref{eqn:example}, the late items are fairly small -- and this is why FCFS is near-optimal. 
When inventory is small then anything that FCFS successfully fits into the knapsack is already very large, relative to the small capacity.
So FCFS has a near-optimal performance when inventory is very small.
In the moderate inventory regime, FCFS has a bad performance.
This is because, when inventory is around $100$, accepting the first two (small) items will block the third (large) item. 
Whenever there is a large item that arrives after a small item, such as the arrival instance \eqref{eqn:example} highlighted by this paper, the performance of FCFS will be bad.
In some sense, the arrival instance \eqref{eqn:example} is a bad instance for FCFS. As we will see in the next simulation, FCFS has a much better performance in the average case.
Our proposed algorithm has a relatively smaller variance compared with FCFS, and seems to be reasonable uniformly.
Finally, the algorithm in HKM15 seems to have a performance that is always just over $50\%$, even in the small inventory or large inventory regimes, due to its artificially saving inventory with $50\%$ probability.

\subsection{Average and Worst Case Performance over all SKUs}
The earlier results shown for a specific SKU was used to illustrate the experimental setup. 
Now we show aggregate experimental results over all the SKUs.
There are $974$ different SKUs, carried in $21$ different warehouses over the country.
For any fixed scaling factor, we compute the average performance over different SKUs.

We implement our \textbf{Random-Threshold} algorithm in the following three ways.
The first implementation involves independently generating a random threshold for each SKU and for each warehouse.
The second implementation involves generating evenly-spaced percentiles of the threshold distribution $F$ across different warehouses.
For the $21$ different warehouses, we take 21 evenly-spaced percentiles\footnote{Note that some SKUs are only stored in a subset of them, say, only 6 warehouses. And we take 6 evenly-spaced percentiles.} of the threshold distribution $F$, that is, we take the 21 thresholds defined by $F^{-1}(0), F^{-1}(\frac{1}{21}), ..., F^{-1}(\frac{20}{21})$.
Then we randomly permute them and assign them over the 21 warehouses, making accept/reject decisions at each warehouse based on the assigned threshold (scaled by the starting inventory).
The third implementation involves independently generating one single percentile for each SKU, and use this percentile for all warehouses.
These three implementations all have very similar performance.
These three implementations are referred to as Random-Threshold 1, Random-Threshold 2, and Random-Threshold 3.

\begin{figure}[t]
\centering
\includegraphics[width=0.95\linewidth]{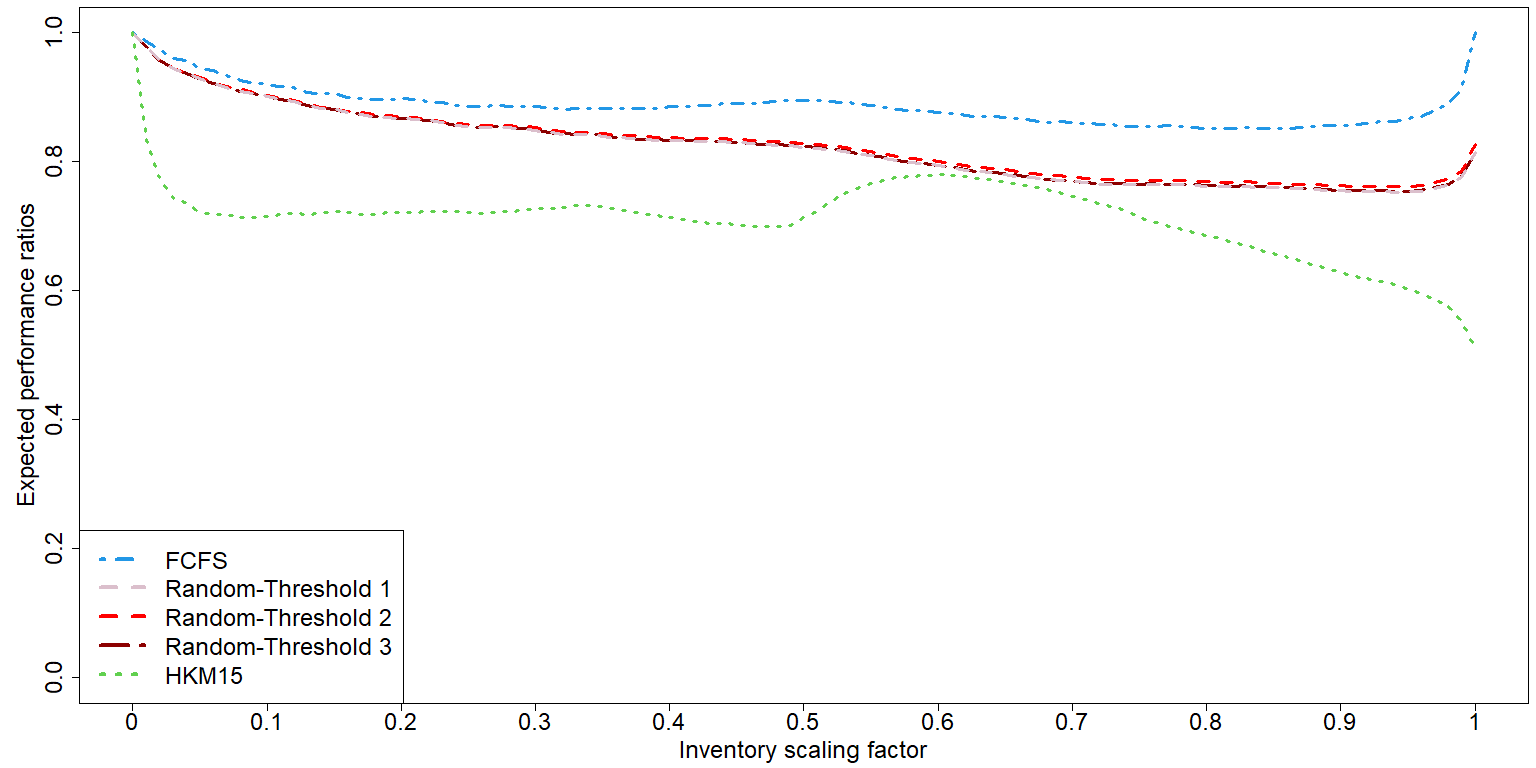} 
\caption{Computational of the average case performance using real arrival sequences over the country.}
\label{fig:computational2}
\end{figure}

We compare our performance against the (deterministic) revenue from the greedy first-come-first-serve policy, named \textbf{FCFS} and the expected revenue from the algorithm suggested in \citet{han2015randomized}, named \textbf{HKM15}.
The results are shown in Figure~\ref{fig:computational2}, where we have divided all the numbers by its corresponding offline optimal integer packing.
We find that the FCFS policy has the best average-case performance ratio.
While this is discouraging, we believe that the way in which order sizes are censored in our data favors FCFS, since large orders cannot come at the end.
Still, the gap between FCFS and our Random-Threshold algorithm is always smaller than $7\%$, indicating that the performance of our Random-Threshold algorithm is not much worse than FCFS.
If we compare our Random-Threshold algorithm with HKM15, we see that our Random-Threshold algorithm \textit{always} outperforms HKM15.
The gap between these two algorithms is between $5\%$--$27\%$.

\begin{figure}[t]
\centering
\includegraphics[width=0.95\linewidth]{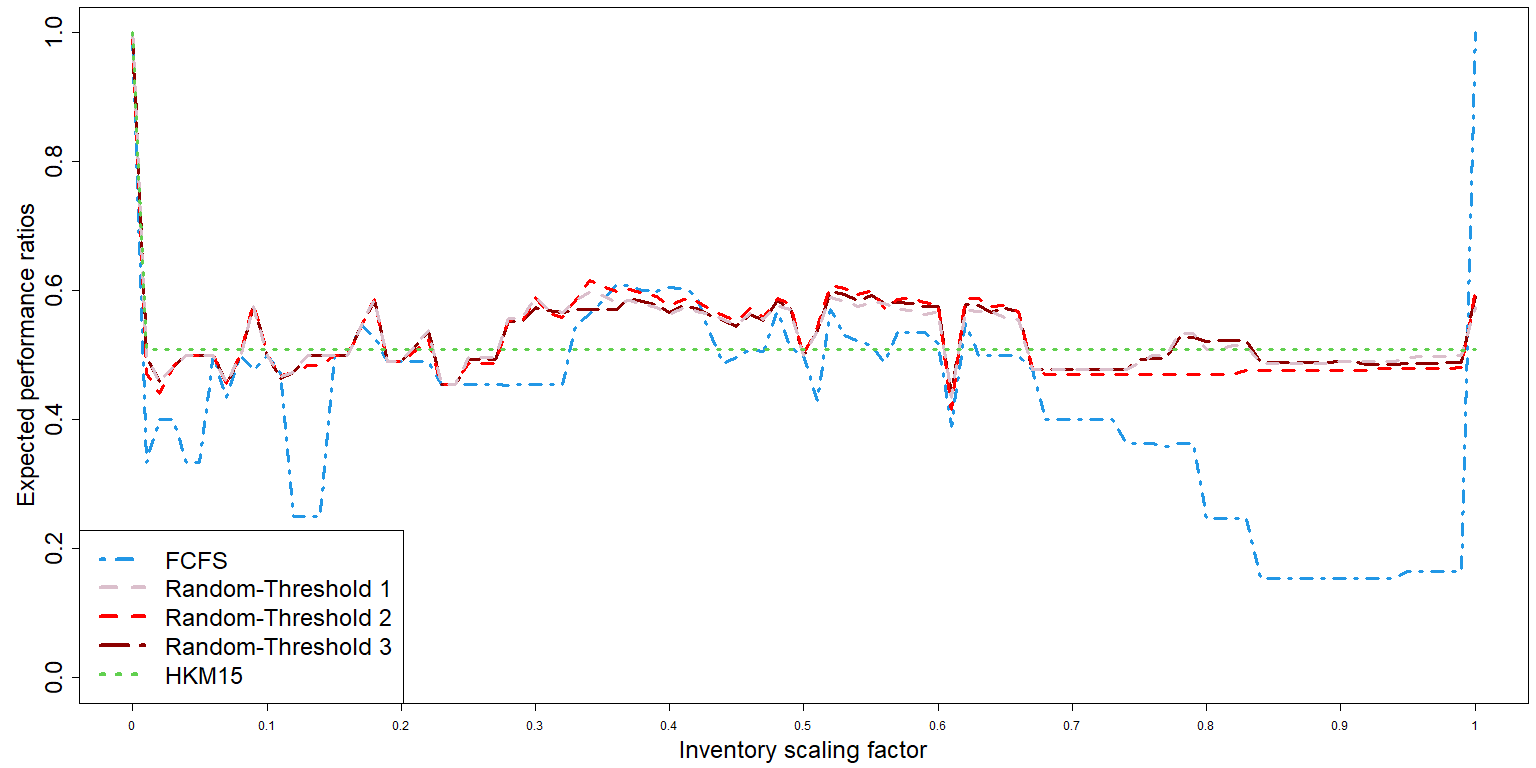} 
\caption{Computational of the worst case performance using real arrival sequences over the country.}
\label{fig:computational3}
\end{figure}

Finally, to illustrate the benefit of our Random-Threshold algorithm, we also present the \textit{worst case} performance of each algorithm over the SKUs, for each scaling factor.
The computational results are shown in Figure~\ref{fig:computational3}.
We see that when inventory is very small, FCFS, HKM15, and our Random-Threshold algorithms all have near-optimal performance.
Our Random-Threshold algorithms have the best worst case performance for a large fraction of scaling factors.
This shows that setting different thresholds at different warehouses, according to the distributions we derived, indeed provides the best baseline guarantee on the fulfillment of any SKU.
Note that the performance of FCFS does not have any guarantee -- FCFS can have a performance as bad as 15\%.
Meanwhile, the performance for HKM15 always equals its theoretical guarantee of $50\%$.

\section{Conclusion}

In this paper, we study a wholesale supply chain ordering problem, and model it as an online unit-density knapsack problem.
The wholesale supply chain ordering problem, which imposes the knapsack constraint, is fundamentally different from a regular supply chain ordering problem. 
As we show in Theorem~\ref{thm:knapsackDEexample}, there is a separation between the upper bound (impossibility result) under the wholesale supply chain problem and the lower bound under the regular supply chain problem.
Therefore, the wholesale supply chain ordering problem deserves special attention to design new algorithms.

We focus on a particular class of random threshold algorithms, that initially draw a random threshold and never change the threshold throughout the entire horizon.
This class of algorithms can be naturally applied to the supply chain ordering practice, as we demonstrated using data from a Latin American chain department store.
The threshold policy has many benefits such as explainability, simplicity, and incentive-compatibility. 
Even though such benefits are difficult to quantify, we can quantify the cost of restricting ourselves to threshold policies by a small gap in competitive ratios. 
Our results support industry practitioners to adopt threshold policies.

Finally, we conclude our paper by pointing out two possible extensions.
First, we could extend our model to consider multiple consumptions over different knapsacks, which, instead of packing numbers, the decision maker packs vectors (see, e.g., \citet{azar2013tight}).
Suppose we are managing a manufacturing plant that requires different resources to make products, instead of a warehouse managing a single stock.
One arriving item could potentially require more than one single resource to be produced.
For each unit of resource consumed, there is an associated revenue / cost.
Our goal is to maximize the total revenue / minimize the total cost throughout the horizon.
Second, we could also extend our model to reusable products.
Suppose there is a fixed total amount of cloud computing resources whose capacity is 1, and an unknown sequence of tasks with sizes at most 1.
The tasks arrive one-by-one, and each task must be irrevocably either assigned to some amount of computing resources, or immediately discarded upon arrival.
A task can be accepted only if its requested resources do not exceed the remaining resource capacity.
Associated with each accepted task, there is a usage time after which the computing resources can be returned.
These resources immediately become available after the usage time.
If one unit of resource is occupied for one period of time, a constant amount of revenue is generated.
Our goal is to maximize the total revenue generated throughout the horizon.
Strong assumptions of the throughput might be made to show non-trivial competitive ratios.

\ACKNOWLEDGMENT{The authors thank \L ukasz Je\.{z} for pointing us to several references, including \citet{han2015randomized} and \citet{cygan2016online}. The authors thank the department editor George Shanthikumar, the anonymous AE, and the anonymous reviewers for their insightful suggestions. In particular, Theorem~\ref{thm:generalization} is shown by one of our reviewers.}

\begingroup
\OneAndAHalfSpacedXI
\bibliographystyle{informs2014} 
\bibliography{bibliography} 
\endgroup

\ECSwitch


\ECHead{E-Companion}

\section{Proof of Theorem~\ref{thm:3/7}}
\label{sec:proof:thm:3/7}

\proof{Proof of Theorem~\ref{thm:3/7}.}
For any instance of arrival sequence $S$, we will show $\frac{\ALGN(S)}{\OPTp(S)} \geq 3/7$.

First of all, $\Gre$ always accepts at least one item. Denote the set of items accepted by $\Gre$ as $G$, and we know $G \ne \emptyset$. Denote $\size(G) = g$. If $G=[T]$ then $\Gre$ is optimal. In this case $$\frac{\ALGN}{\OPTp} \geq \Pr(\tau=0) \cdot 1 + \Pr(\tau > 0) \cdot 0 \geq F(0) = 4/7 \geq 3/7.$$

If $G \subsetneqq [T]$, let $M=[T] \backslash G$ denote the set of items blocked by $\Gre$. Since $\Gre$ always accepts an item as long as it can fill in, any item blocked by $\Gre$ must exceed the remaining space of the knapsack, at the moment it is blocked.

\begin{figure}[!h]\centering
\caption{Illustration of the items that $\Gre$ accepts, and blocks. Note that $m$ is the size of the smallest item in $M$, not necessarily the first.}
\label{fig:mainillustrator}
\includegraphics[width=0.7\textwidth,]{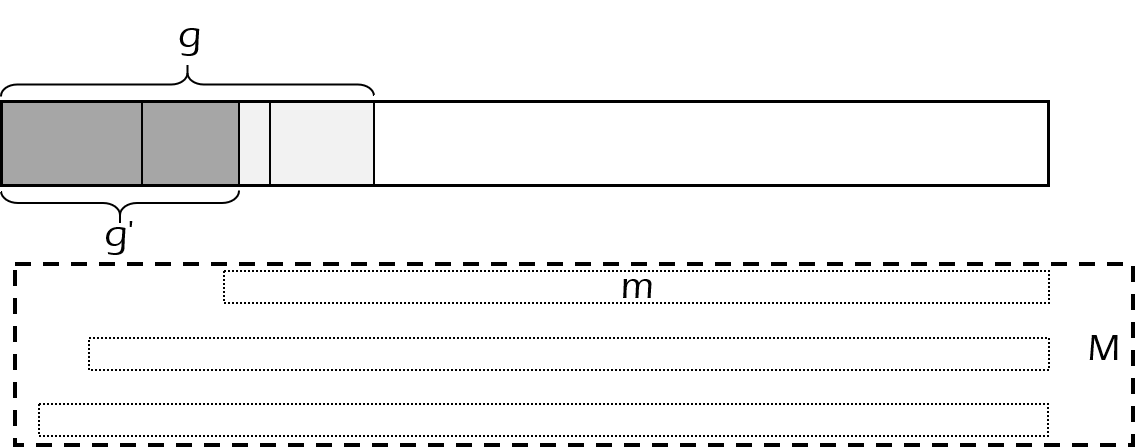}
\end{figure}

Let $m$ be the size of the smallest item in $M$, i.e. $m = \min_{t \in M} s_t$. Define index $t_m$ for the smallest item, or the first smallest item, if there are multiple smallest items,
\begin{align}
t_m = \min \left\{ t\in[T] \left| s_t = m \right. \right\}. \label{eqn:tildet}
\end{align}
Denote $G'$ as the set of items accepted by $\Gre$, at the moment item $t_m$ is blocked.
Let $g'= \size(G')$. See Figure~\ref{fig:mainillustrator}. A straightforward, but useful observation about $m$ is:
\begin{equation}
\label{eqn:gandm}
g'+m>1,
\end{equation}
because $m$ is blocked by $\Gre$.

Next, we wish to understand when we can pack an item of size at least $m$, by selecting a proper threshold $\tau$.
We distinguish two cases: $m \geq 1/2$ and $m<1/2$.

\noindent\textbf{Case 1}: $m \geq 1/2$.\\
Let $S^{\THR}(\tau)$ be the set of items that have sizes at least $\tau$, i.e. $S^{\THR}(\tau) = \left\{t \in S \left| s_t \geq \tau \right.\right\}$. Now define
\begin{equation}
\label{eqn:defineq}
\begin{split}
q = \max & \quad \tau\\
s.t. & \quad m+ \size( S^{\THR}(\tau) \cap G' ) > 1.
\end{split}
\end{equation}
This means that if we adopt a $\THR(q)$ policy, then item $t_m$ must not be accepted because $m+ \size( S^{\THR}(\tau) \cap G' ) > 1$.
That is, the items that a $\THR(q)$ policy accepts before item $t_m$ arrives would exceed $1-m$, leaving not enough space to accept item $t_m$ the size $m$ item\footnote{This does not exclude the possibility that it is also rejected, due to $q>m$, which leads to the discussion in Case 1.1.}. 
\begin{figure}[!h]\centering
\caption{Illustration of Case 1 (and specifically, Case 1.2)}
\label{fig:case1}
\includegraphics[width=0.7\textwidth,]{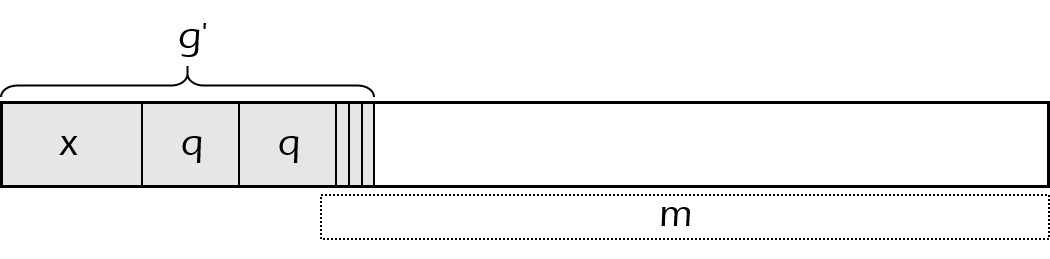}
\end{figure}

Now consider the items in $S^{\THR}(q) \cap G'$. These items have sizes at least $q$. We count how many size $q$ items are there, and let $n$ be the number of size $q$ items. Denote the total size of the remaining items in $S^{\THR}(q) \cap G'$ be $x$. We know that $\size(S^{\THR}(q) \cap G') = nq+x$. See Figure~\ref{fig:case1}.

We make the following observations:
\begin{enumerate}
\item There must exist some item from $G'$ that is of size $q$, i.e.
\begin{equation}
\label{eqn:qexistance}
\exists t_q \in G' \subseteq [T], s.t. \ s_{t_q} = q.
\end{equation}
This is because otherwise we can select the smallest item in $G'$ whose size is (strictly) larger than $q$. This size satisfies \eqref{eqn:defineq}, and violates the maximum property of $q$.
\item Size $m$ items can not fit in together with all the items in $S^{\THR}(q) \cap G'$, i.e., 
\begin{equation}
\label{eqn:qandxandm}
nq+x+m>1
\end{equation}
This is because $\size(S^{\THR}(q) \cap G') = nq+x$. And then the constraint in \eqref{eqn:defineq} implies \eqref{eqn:qandxandm}.
\item A size $m$ item can fit in together with items $S^{\THR}(\tau) \cap G', \forall \tau > q$, i.e., 
\begin{equation}
\label{eqn:xandm}
x+m \leq 1
\end{equation}
This is because otherwise $x+m>1$, and we could further strictly increase $q$ to still satisfy the constraint in \eqref{eqn:defineq}. Define $$\hat{q} = \min \left\{s_t \left| s_t > q, t \in S^{\THR}(q) \cap G' \right.\right\}.$$ We know (i) $\hat{q} > q$; (ii) $\size(S^{\THR}(\hat{q}) \cap G') + m = x + m > 1$. So $\hat{q}$ violates the maximum property of $q$.
\end{enumerate}

We further distinguish two cases: $q > m$, and $q \leq m$.

\noindent\textbf{Case 1.1}: $q > m$.\\
In this case, if we adopt $\Gre$ then we can get $g$.

If we adopt $\THR(\tau), \forall \tau \in (0,q]$ then we can get no less than $q$. This is because due to \eqref{eqn:qexistance} there must exist some item $t_q \in G'$ of size $q$. We either accept it, in which case we immediately earn $q$, or we block it because we have accepted some item $z\in[T]$ from $M$ that arrived earlier and consumed too much space. In the latter case, $\Gre$ blocks item $z$ earlier than it accepts item $t_q$, which means that $s_z$ is no less than the remaining capacity when $z$ arrives, which is no less than the remaining capacity when $t_q$ arrives at a later time, which is further no less than $q$, the size of $t_q$ which is accepted. So we have $s_z \geq s_{t_q} = q$ in the second case. Putting both cases together we earn $q$.

We have the following:
\begin{align*}
\ALGN & \geq \Pr(\tau=0) \cdot g + \Pr(0 < \tau \leq q) \cdot q \\
& = \FN(0) \cdot g + (\FN(q) - \FN(0)) \cdot q \\
& \geq \FN(0) \cdot (1-2q) + \FN(q) \cdot q \\
& = 4/7 \cdot (1-2q) + 1 \cdot q \\
& = 4/7 - 1/7 \cdot q \\
& \geq 3/7
\end{align*}
where the second inequality is because $g \geq g' > 1-m$ (due to \eqref{eqn:gandm}) and $1-m > 1-q$ (Case 1.1: $q>m$); second equality is because $q > m \geq 1/2$ and the way we defined $\FN(\cdot)$ in \eqref{eqn:defineF} so $\FN(q)=1$; last inequality is because $q \leq 1$.

Since $\OPTp \leq 1$, we have $\frac{\ALGN}{\OPTp} \geq \frac{3}{7}$.

\noindent\textbf{Case 1.2}: $q \leq m$.\\
In this case, if we adopt $\Gre$ then we can get as much as $g$.

If we adopt $\THR(\tau), \forall \tau \in (0,q]$ then we get no less than $q$. This is because due to \eqref{eqn:qexistance} there must exist some item $t_q \in G'$ of size $q$. We either accept it, in which case we immediately earn $q$, or we block it because we have accepted some item $z\in[T]$ from $M$ that arrived earlier and consumed too much space. In the latter case, $\Gre$ blocks item $z$ earlier than it accepts item $t_q$, which means that $s_z \geq s_{t_q} = q$. So in either case we earn $q$.

If we adopt $\THR(\tau), \forall \tau \in (q,m]$ then we get no less than $m$. This is because due to \eqref{eqn:xandm}, all the items from $S^{\THR}(\tau) \cap G'$ altogether will not block item $t_m$ (from expression \eqref{eqn:tildet}); and $\tau \leq m$ so we will not reject item $t_m$. We either accept item $t_m$, in which case we immediately earn $m$, or we block it because we have accepted some item $z\in[T]$ from M and consumed too much space. But $m$ is smallest item size in $M$, which means that $s_z \geq m$. So in either case we earn $m$.

We have the following:
\begin{align*}
\ALGN & \geq \Pr(\tau=0) \cdot g + \Pr(0 < \tau \leq q) \cdot q + \Pr(q < \tau \leq m) \cdot m\\
& = \FN(0) \cdot g + (\FN(q) - \FN(0)) \cdot q + (\FN(m)-\FN(q)) \cdot m\\
& \geq \FN(0) \cdot (nq+x) + (\FN(q) - \FN(0)) \cdot q + (\FN(m) - \FN(q)) \cdot (1-(nq+x))\\
& =  (\FN(q) - \FN(0)) \cdot q + 1 - \FN(q) + (\FN(q) - 3/7) \cdot (nq+x)\\
& \geq (\FN(q) - \FN(0)) \cdot q + 1 - \FN(q) + (\FN(q) - 3/7) \cdot q\\
& = \FN(q) \cdot (2q-1) + 1 - q
\end{align*}
where the second inequality is because $g \geq g' \geq nq+x$ and $m > 1-(nq+x)$ (due to \eqref{eqn:qandxandm}); second equality is because $m \geq 1/2$ and the way we defined $\FN(\cdot)$ in \eqref{eqn:defineF} so $\FN(m)=1$; the last inequality is because $\FN(q) \geq \FN(0) =4/7 > 3/7$, so the coefficient in front of $nq+x$ is positive.

Now we plug in the expression of $\FN(q)$ as defined in \eqref{eqn:defineF}. If $q \leq 3/7$ then $\ALGN \geq \frac{4/7-q}{1-2q} \cdot (2q-1) + 1-q = 3/7;$ If $q > 3/7$ then $\ALGN \geq 1\cdot (2q-1) + 1 - q = q > 3/7.$ So in either case we have shown $\ALGN \geq 3/7$.

Since $\OPTp \leq 1$, we have $\frac{\ALGN}{\OPTp} \geq \frac{3}{7}$.

\begin{figure}[!htb]\centering
\includegraphics[width=0.7\textwidth,]{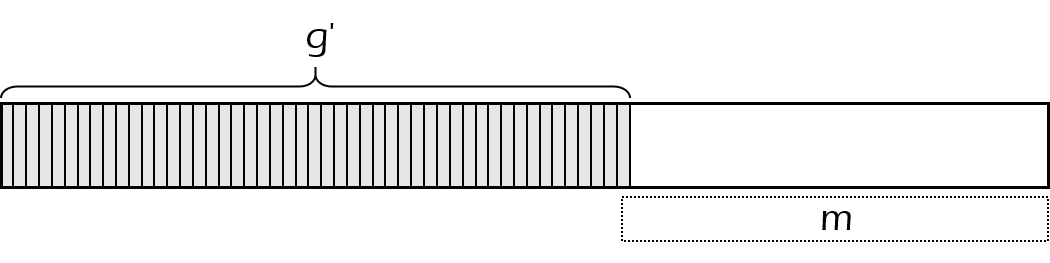}
\caption{Illustration of Case 2}
\label{fig:case2}
\end{figure}

\noindent\textbf{Case 2}: $m < 1/2$.\\
In this case, a crude analysis is enough.
See Figure~\ref{fig:case2}.

If we adopt $\Gre$ then we can get as much as $g$. 

If we adopt $\THR(\tau), \forall \tau \in (0,m]$ then we either get $m$, or $m$ is blocked, in which case we must have already earned at least $1-m$ to block $m$.

We have the following:
\begin{align*}
\ALGN & \geq \Pr(\tau=0) \cdot g + \Pr(0 < \tau \leq m) \cdot \min\{m,1-m\} \\
& \geq \Pr(\tau=0) \cdot g + \Pr(0 < \tau \leq m) \cdot m \\
& = \FN(0) \cdot g + (\FN(m) - \FN(0)) \cdot m \\
& \geq \FN(0) \cdot (1-m) + (\FN(m) - \FN(0)) \cdot m \\
& = \FN(m) \cdot m + 4/7 \cdot (1-2m)
\end{align*}
where the second inequality is because $m<1/2$; the last inequality is because $g \geq g' > 1-m$ (due to \eqref{eqn:gandm}).

Now we plug in the expression of $\FN(m)$ as defined in \eqref{eqn:defineF}. If $m > 3/7$ then 
\begin{align*}
\ALGN \geq 4/7 - 1/7 \cdot m \geq 3/7,
\end{align*}
because $m < 1/2 \leq 1$.
If $m \leq 3/7$ then 
\begin{align*}
\ALGN \geq \frac{4/7-m}{1-2m} \cdot m + \frac{4}{7} \cdot (1-2m).
\end{align*} 
Note that
\begin{align*}
\frac{4/7-m}{1-2m} \cdot m + \frac{4}{7} \cdot (1-2m) = \frac{9}{28} \cdot (1-2m) + \frac{1}{28} \cdot \frac{1}{1-2m} + \frac{3}{14} \geq 2 \sqrt{\frac{9}{28}\cdot \frac{1}{28}} + \frac{3}{14} = \frac{3}{7}.
\end{align*}
So in either case we have $\ALGN \geq 3/7$.

Since $\OPTp \leq 1$, we have $\frac{\ALGN}{\OPTp} \geq \frac{3}{7}$.

In all, we have enumerated all the possible cases, to find $\dfrac{\ALGN}{\OPTp} \geq \dfrac{3}{7}$ always holds. 
\Halmos\endproof

\section{Proof of Theorem~\ref{thm:0.4324}} \label{sec:0.4324alg:proof}

\proof{Proof of Theorem~\ref{thm:0.4324}.}
We are going to show that, for any instance of arrival sequence $S$, we have $\frac{\ALGG(S)}{\OPT(S)} \geq \cG$. We lower bound $\ALGG(S)$ and upper bound $\OPT(S)$ at the same time.

First of all, if the arrival sequence is not empty set, $\Gre$ always accepts something. Denote the set of items accepted by $\Gre$ as $G$. Denote $\size(G) = g$. If $G=[T]$ then $\Gre$ is optimal. In this case $$\frac{\ALGG}{\OPT} \geq \Pr(\tau=0) \cdot 1 + \Pr(\tau > 0) \cdot 0 \geq \FG(0) = 1-\cG \geq \cG.$$

If $G \subsetneqq [T]$, let $M=[T] \backslash G$ denote the set of items blocked by $\Gre$. Since $\Gre$ always accepts an item as long as it can fill in, any item blocked by $\Gre$ must exceed the remaining space of the knapsack, at the moment it is blocked. We also know that $G \cup M = [T]$, $G \cap M = \emptyset$.

Let $m$ be the smallest size in $M$, i.e. $m = \min_{t \in M} s_t$. Define index $t_m$ for the smallest item, or the first smallest item, if there are multiple smallest items.
\begin{align}
t_m = \min \left\{ t\in[T] \left| s_t = m \right. \right\}. \label{eqn:tildet2}
\end{align}
Denote $G'$ as the set of items accepted by $\Gre$, at the moment $s_{t_m}$ is blocked.
Let $g'= \size(G')$. See Figure~\ref{fig:mainillustrator}. A straightforward, but useful observation about $m$ is:
\begin{equation}
\label{eqn:gandm2}
g'+m>1,
\end{equation}
because $m$ is blocked by $\Gre$.
We wish to understand when we can accept an item of size at least $m$, by selecting a proper threshold $\tau$.

We distinguish two cases: $m > 1/2$ and $m \leq 1/2$.

\noindent\textbf{Case 1}: $m > 1/2$.\\
Let $S^{\THR}(\tau)$ be the set of items that have sizes at least $\tau$, i.e. $S^{\THR}(\tau) = \left\{t \in S \left| s_t \geq \tau \right.\right\}$. Now define 
\begin{equation}
\label{eqn:defineq2}
\begin{split}
q = \max & \quad \tau\\
s.t. & \quad m+ \size( S^{\THR}(\tau) \cap G' ) > 1
\end{split}
\end{equation}
This means that if we adopt a $\THR(q)$ policy, then the size $m$ item must be blocked (possibly it will also be rejected, due to $q>m$).

Now consider the items in $S^{\THR}(q) \cap G'$. See Figure~\ref{fig:case1}. These items have sizes at least $q$. We count how many size $q$ items are there, and let $n$ be the number of size $q$ items. Denote the total size of the remaining items by $x$. We know that
\begin{equation}
\label{eqn:nqxexistance2}
\size(S^{\THR}(q) \cap G') = nq+x.
\end{equation}

We make the following observations:
\begin{enumerate}
\item There must exist some item from $G'$ that is of size $q$, i.e.
\begin{equation}
\label{eqn:qexistance2}
\exists t_q \in G' \subseteq [T], s.t. \ s_{t_q} = q.
\end{equation}
This is because otherwise we can select the smallest item size in $G'$ that is also larger than $q$. This item size satisfies \eqref{eqn:defineq2}, and violates the maximum property of $q$.
\item Size $m$ items can not fit in together with items $S^{\THR}(q) \cap G'$, i.e.
\begin{equation}
\label{eqn:qandxandm2}
nq+x+m>1
\end{equation}
This is because $\size(S^{\THR}(q) \cap G') = nq+x$. This is implied by \eqref{eqn:defineq2}.
\item A size $m$ item can fit in together with items $S^{\THR}(\tau) \cap G', \forall \tau > q$, i.e.
\begin{equation}
\label{eqn:xandm2}
x+m \leq 1
\end{equation}
This is because otherwise we could further increase $q$ to $\hat{q}$ so that $\size(S^{\THR}(\hat{q}) \cap G') + m > 1$, which violates the maximum property of $q$.
\end{enumerate}

We further distinguish two cases: $q > m$, and $q \leq m$.

\noindent\textbf{Case 1.1}: $q > m$.\\
In this case, if we adopt $\Gre$ then we can get as much as $g$. 

If we adopt $\THR(\tau), \forall \tau \in (0,q]$ then we can get no less than $q$. This is because due to \eqref{eqn:qexistance2} there must exist some item $t_q \in G'$ of size $q$. We either accept it, in which case we immediately earn $q$, or we block it because we have accepted some item $z\in[T]$ from $M$ that arrived earlier and consumed too much space. In the latter case, $\Gre$ blocks item $z$ earlier than it accepts item $t_q$, which means that $s_z \geq s_{t_q} = q$. So in either case we earn $q$.

We have the following:
\begin{align*}
\ALGG & \geq \Pr(\tau=0) \cdot g + \Pr(0 < \tau \leq q) \cdot q \\
& = \FG(0) \cdot g + (\FG(q) - \FG(0)) \cdot q \\
& \geq \FG(0) \cdot (1-2q) + \FG(q) \cdot q \\
& = (1-\cG) \cdot (1-2q) + \left[2(1-\cG) - \frac{1-2\cG}{q} \right] \cdot q \\
& = \cG
\end{align*}
where the second inequality is because $g \geq g' > 1-m$ (due to \eqref{eqn:gandm2}) and $1-m > 1-q$ (Case 1.1: $q>m$); second equality is because $q > m \geq 1/2 > \qG$, so we plug in $\FG(\cdot)$ as defined in \eqref{eqn:defineF2}.

Since $\OPT \leq 1$, we have $\frac{\ALG}{\OPT} \geq \cG$.

\noindent\textbf{Case 1.2}: $q \leq m$.\\
First we wish to upper bound $\OPT$. $\OPT$ selects some items from $[T]=G \cup M$, where $G \cap M = \emptyset$. Notice that $m > 1/2$ so there is at most $1$ item from $M$ that $\OPT$ can select. If $\OPT$ selects no item from $M$, then $\OPT \leq g$. With probability $\FG(0)$, $\ALGG$ adopts $\Gre$ and earns $g$. So we have $$\frac{\ALGG}{\OPT} \geq \Pr(\tau=0) \cdot 1 + \Pr(\tau > 0) \cdot 0 \geq \FG(0) = 1-\cG \geq \cG.$$

\begin{figure}[!htb]\centering
\includegraphics[width=0.7\textwidth,]{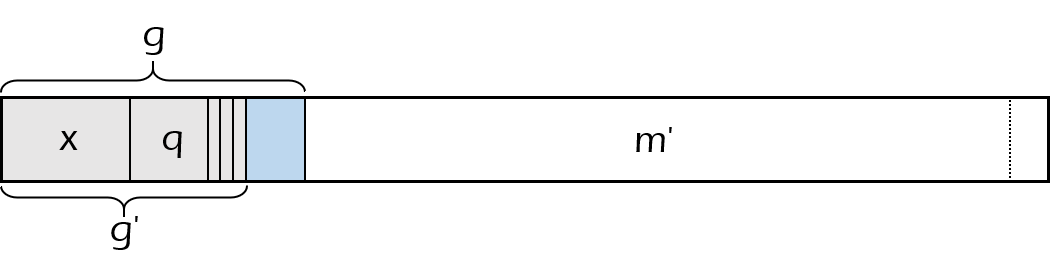}
\caption{Illustration of the items accepted by $\OPT$}
\label{fig:opt}
\end{figure}

If $\OPT$ selects one item from $M$, let $t_{m'} \in [T]$ be this item. So $s_{t_{m'}} = m' \geq m$. See Figure~\ref{fig:opt}.

We can partition all the items in $S$ into three sets:
\begin{align*}
M; && S^\THR (q) \cap G'; && G \backslash (S^\THR (q) \cap G')
\end{align*}
Let $\tilde{g} = \size( G \backslash (S^\THR (q) \cap G') )$. Since $S^\THR (q) \cap G'$ and $G \backslash (S^\THR (q) \cap G')$ form a partition of $G$, we have $g = (nq+x) + \tilde{g}$. From \eqref{eqn:qandxandm2} we know that $m'+\size( S^\THR (q) \cap G' ) \geq m+\size( S^\THR (q) \cap G' ) > 1$. This means that $\OPT$ cannot pack item $t_{m'}$ and $S^\THR (q) \cap G'$ together. $\OPT$ must block at least one item from $\{t_{m'}\} \cup (S^\THR (q) \cap G')$ -- and the smallest item from this union is of size $q$ (because $q \leq m \leq m'$). So we upper bound $\OPT$ by:
\begin{equation}
\label{eqn:upperboundopt}
\begin{split}
\OPT \leq & \min\left\{1, \left[m'+\size( S^\THR (q) \cap G' )\right] - q + \size( G \backslash (S^\THR (q) \cap G') ) \right\}\\
= & \min\left\{1, m'+ (nq+x) - q + \tilde{g} \right\}.
\end{split}
\end{equation}
In other words, we allow $\OPT$ to accept the items of size $\tilde{g}$; but we need to argue that $\OPT$ must block at least $q$ from $G'$.

Then we analyze $\ALGG$. If we adopt $\Gre$ then we can get as much as $g$. 

If we adopt $\THR(\tau), \forall \tau \in (0,q]$ then we get no less than $nq+x$. This is because due to \eqref{eqn:nqxexistance2} there must exist some items in $S^{\THR}(q) \cap G'$, which are of size $nq+x$. For any item in $(S^{\THR}(q) \cap G')$, if we accept them all, we immediately earn $(nq+x)$. Suppose we block some item in $(S^{\THR}(q) \cap G')$ because we have accepted some item $z \in[T]$ from $M$ and consumed too much space. But $\Gre$ blocks item $z$ earlier than it accepts the item in $(S^{\THR}(q) \cap G')$, whose smallest possible size is $q$. This suggests that $s_z \geq q$. But this contradicts with the fact that we do not accept item $z$ using the $\THR(\tau)$ threshold policy when $\tau \leq q$.

If we adopt $\THR(\tau), \forall \tau \in (q,m]$ then we get no less than $m$. This is because due to \eqref{eqn:xandm2}, any item in $S^{\THR}(\tau) \cap G'$ will not block item $t_m$ (from expression \eqref{eqn:tildet2}); and $\tau \leq m$ so we will not reject item $t_m$. We either accept item $t_m$, in which case we immediately earn $m$, or we block it because we have accepted some item $z\in[T]$ from M and consumed too much space. But $m$ is smallest item size in $M$, which means that $s_z \geq m$. So in either case we earn $m$.

If we adopt $\THR(\tau), \forall \tau \in (m,m']$ then we get no less than $\tau$. This is because $s_{t_{m'}}$ does exist, and $\THR(\tau)$ must accept at least one item. The least that $\THR(\tau)$ can get is $\tau$.

We have the following:
\begin{align*}
\ALGG & \geq \Pr(\tau=0) \cdot g + \Pr(0 < \tau \leq q) \cdot (nq+x) + \Pr(q < \tau \leq m) \cdot m + \int_m^{m'} \tau \ \mathrm{d} \FG(\tau)\\
& = \FG(0) \cdot (nq+x + \tilde{g}) + (\FG(q) - \FG(0)) \cdot (nq+x) + (\FG(m)-\FG(q)) \cdot m + \int_m^{m'} \tau \ \mathrm{d} \FG(\tau) \\
& = \FG(0) \cdot \tilde{g} + \FG(q) \cdot (nq+x-m) + \FG(m') \cdot m' - \int_m^{m'} \FG(\tau) \ \mathrm{d} \tau \\
& = \FG(0) \cdot \tilde{g} + \FG(q) \cdot (nq+x-m) + \FG(m') \cdot m' - \int_m^{m'} \left(2(1-\cG) - \frac{1-2\cG}{\tau}\right) \ \mathrm{d} \tau \\
& \geq \FG(0) \cdot \tilde{g} + \FG(q) \cdot (2(nq+x)-1) + \FG(m') \cdot m' - \int_{1-(nq+x)}^{m'} \left(2(1-\cG) - \frac{1-2\cG}{\tau}\right) \ \mathrm{d} \tau
\end{align*}
where the second equality is due to integration by part (our definition of $\FG(\cdot)$ in \eqref{eqn:defineF2} is a continuous function); the third equality is because $m>1/2 > \qG$ and we plug in the expression of $\FG(\cdot)$; the last inequality is because $\frac{\partial \ALGG }{\partial m} = \FG(m) - \FG(q) \geq 0$, (because $q \leq m$, and $\FG(\cdot)$ is a increasing function), so that $\ALGG$ is increasing in $m$. Hence, $\ALGG$ achieves its minimum when $m$ is the smallest, and $m > 1 - (nq+x)$ from \eqref{eqn:qandxandm2}.

Observe that
\begin{multline*}
\ALGG - \cG \OPT \geq \FG(0) \cdot \tilde{g} + \FG(q) \cdot (2(nq+x)-1) + \FG(m') \cdot m' \\
- \int_{1-(nq+x)}^{m'} \left(2(1-\cG) - \frac{1-2\cG}{\tau}\right) \ \mathrm{d} \tau - \cG \cdot \min\left\{1, m'+ (nq+x) - q + \tilde{g} \right\}
\end{multline*}
If we focus on the dependence of $\tilde{g}$, we find that $$\frac{\partial \left( \ALGG - \cG \OPT \right)}{\partial \tilde{g}} \geq \FG(0) - \cG = 1 - 2\cG \geq 0,$$ where the first inequality is because the subgradient of the subtracted term is either $0$ or $\cG$. Since $\ALGG - \cG \OPT$ is a increasing function of $\tilde{g}$, it achieves its minimum when $\tilde{g} = 0$.

Thus,
\begin{multline*}
\ALGG - \cG \OPT \geq \FG(q) \cdot (2(nq+x)-1) + \FG(m') \cdot m' \\
- \int_{1-(nq+x)}^{m'} \left(2(1-\cG) - \frac{1-2\cG}{\tau}\right) \ \mathrm{d} \tau - \cG \cdot \min\left\{1, m'+ (nq+x) - q \right\}
\end{multline*}

Now let $y = (n-1)q + x$, and we plug in $\FG(\cdot)$ as we defined in \eqref{eqn:qandxandm2}.

\noindent\textbf{Case 1.2.1}: When $q \leq \qG$, we have:
\begin{align*}
& \ALGG - \cG \OPT \\
\geq & \FG(q) \cdot (2(q+y)-1) + \FG(m') \cdot m' - \int_{1-(q+y)}^{m'} \left(2(1-\cG) - \frac{1-2\cG}{\tau}\right) \ \mathrm{d} \tau - \cG \cdot \min\left\{1, m'+ y \right\} \\
= & (1-\cG)(2(q+y) - 1) + \frac{(2q-1+2y)(1-2\cG)\ln{(1-q)}}{2q-1} + 2(1-\cG)m' - (1-2\cG)\\
& -2(1-\cG)m' + 2(1-\cG)[1-(q+y)] + (1-2\cG)\left[ \ln{m'} - \ln{(1-(q+y))} \right] - \cG \cdot \min\left\{1, m'+ y \right\}\\
= & \cG + \frac{(2q-1+2y)(1-2\cG)\ln{(1-q)}}{2q-1} + (1-2\cG)\left[ \ln{m'} - \ln{(1-(q+y))} \right] - \cG \cdot \min\left\{1, m'+ y \right\}
\end{align*}
If we focus on the dependence of $m'$, we will see that $\ALGG - \cG \OPT$ has only one local minimum: when $m' < 1-y$ we have $$\frac{\partial \left( \ALGG - \cG \OPT \right)}{\partial m'} = \frac{1-2\cG}{m'} - \cG \leq \frac{1-2\cG}{1/2} - \cG = 2 - 5\cG <0,$$ because $m' \geq m \geq 1/2$. So $\ALGG - \cG \OPT$ is decreasing on $m'$ when $m' < 1-y$. When $m' > 1-y$ we have $$\frac{\partial \left( \ALGG - \cG \OPT \right)}{\partial m'} = \frac{1-2\cG}{m'} > 0,$$ so $\ALGG - \cG \OPT$ is increasing on $m'$. Hence, $\ALGG - \cG \OPT$ achieves its minimum when $m' = 1-y$.

Plugging in $m' = 1-y$, we have further
\begin{align*}
\ALGG - \cG \OPT \geq & \frac{(2q-1+2y)(1-2\cG)\ln{(1-q)}}{2q-1} + (1-2\cG)\left[ \ln{(1-y)} - \ln{(1-(q+y))} \right]
\end{align*}
If we focus on the dependence of $y$, we find that $$\frac{\partial \left( \ALGG - \cG \OPT \right)}{\partial y} = (1-2\cG)\left[ 2\frac{\ln{(1-q)}}{2q-1} - \frac{1}{1-y} + \frac{1}{1-y-q} \right]>0,$$ because $\ln{(1-q)} < 0, 2q-1 < 2\qG - 1 < 0, \frac{1}{1-y-q} - \frac{1}{1-y} \geq 0$. Since $\ALGG - \cG \OPT$ is increasing on $y$, it achieves its minimum when $y=0$.

Finally, plugging in $y=0$, we have
\begin{align*}
\ALGG - \cG \OPT \geq & \frac{(2q-1)(1-2\cG)\ln{(1-q)}}{2q-1} - (1-2\cG) \ln{(1-q)} = 0
\end{align*}

\noindent\textbf{Case 1.2.2}: When $q > \qG$, we have:
\begin{align*}
& \ALGG - \cG \OPT \\
\geq & \FG(q) \cdot (2(q+y)-1) + \FG(m') \cdot m' - \int_{1-(q+y)}^{m'} \FG(\tau) \ \mathrm{d} \tau - \cG \cdot \min\left\{1, m'+ y \right\} \\
= & 2(1-\cG)(2(q+y) - 1) - \frac{(1-2\cG)(2(q+y) - 1)}{q} + 2(1-\cG)m' - (1-2\cG)\\
& -2(1-\cG)m' + 2(1-\cG)[1-(q+y)] + (1-2\cG)\left[ \ln{m'} - \ln{(1-(q+y))} \right] - \cG \cdot \min\left\{1, m'+ y \right\}\\
= & 2(1-\cG)(y+q) - (1-2\cG) - \frac{(1-2\cG)(2(q+y) - 1)}{q} \\
& + (1-2\cG)\left[ \ln{m'} - \ln{(1-(q+y))} \right] - \cG \cdot \min\left\{1, m'+ y \right\}
\end{align*}
Again, if we focus on the dependence of $m'$, we will see that $\ALGG - \cG \OPT$ has only one local minimum when $m' = 1-y$.

Plugging in $m' = 1-y$, we have further
\begin{multline*}
\ALGG - \cG \OPT \geq \\
(1-\cG)(2(y+q) -1) - \frac{(1-2\cG)(2(q+y) - 1)}{q} + (1-2\cG)\left[ \ln{(1-y)} - \ln{(1-(q+y))} \right]
\end{multline*}
Again, if we focus on the dependence of $y$, we find that $$\frac{\partial \left( \ALGG - \cG \OPT \right)}{\partial y} = 2(1-\cG-\frac{1-2\cG}{q})+ (1-2\cG)\left[ - \frac{1}{1-y} + \frac{1}{1-y-q} \right]>0,$$ because $1-\cG-\frac{1-2\cG}{q} \geq 1-\cG-\frac{1-2\cG}{\qG} \approx 0.142 > 0, \frac{1}{1-y-q} - \frac{1}{1-y} \geq 0$. Since $\ALGG - \cG \OPT$ is increasing on $y$, it achieves its minimum when $y=0$.

Finally, plugging in $y=0$, we have
\begin{align*}
\ALGG - \cG \OPT \geq & (1-\cG)(2q -1) - \frac{(1-2\cG)(2q - 1)}{q} - (1-2\cG) \ln{(1-q)} \\
= & (2q-1)\left[ (1-\cG) - \frac{(1-2\cG)}{q} \right] - (1-2\cG) \ln{(1-q)} \\
\geq & (2q-1)\left[ \frac{(1-2\cG) \cdot (-\ln{(1-q)})}{1-2q} \right] - (1-2\cG) \ln{(1-q)} \\
= & 0
\end{align*}
where the second inequality is because $H(\cG,q) = \frac{1-2\cG}{q} - \frac{(1-2\cG)\ln{(1-q)}}{1-2q} - (1-\cG) \geq 0, \forall q \in (0,1/2)$ from \eqref{eqn:defineHproperty1}, and when $q \in [1/2,1]$, the second line expression is an increasing function of $q$ (because $2q-1; (1-\cG) - \frac{(1-2\cG)}{q};$ and $- (1-2\cG) \ln{(1-q)}$ are all increasing in $q$), thus plugging in $q=1/2$ we have $\ALGG - \cG \OPT \geq - (1-2\cG) \ln{(1-q)} > 0$.

In all, $\ALGG \geq \cG \OPT$.

\noindent\textbf{Case 2}: $m \leq 1/2$.\\
In this case, a crude analysis is enough. See Figure~\ref{fig:case2}.

If we adopt $\Gre$ then we can get as much as $g$. 

If we adopt $\THR(\tau), \forall \tau \in (0,m]$ then we either get $m$, or $m$ is blocked, in which case we must have already earned at least $1-m$ to block $m$.

We have the following:
\begin{align*}
\ALG & \geq \Pr(\tau=0) \cdot g + \Pr(0 < \tau \leq m) \cdot \min\{m,1-m\} \\
& \geq \Pr(\tau=0) \cdot g + \Pr(0 < \tau \leq m) \cdot m \\
& = \FG(0) \cdot g + (\FG(m) - \FG(0)) \cdot m \\
& \geq \FG(0) \cdot (1-m) + (\FG(m) - \FG(0)) \cdot m \\
& = \FG(0) \cdot (1-2m) + \FG(m) \cdot m \\
& \geq (1-\cG)(1-2m) + \left[ 2(1-\cG) - \frac{1-2\cG}{m} \right] \cdot m \\
& = \cG
\end{align*}
where the second inequality is because $m\leq1/2$; the third inequality is because $g \geq g' > 1-m$ (due to \eqref{eqn:gandm2}); the last inequality is because we plug in $\FG(\cdot)$ as defined in \eqref{eqn:defineF2}, and using the fact that $(1-\cG) - \frac{(1-2\cG) \ln{(1-x)}}{1-2x} > 2(1-\cG) - \frac{1-2\cG}{x}$ (the first piece is larger than the second piece) in \eqref{eqn:defineF2}.

Since $\OPT \leq 1$, we have $\frac{\ALG}{\OPT} \geq \cG$.

In all, we have enumerated all the possible cases, to find $\dfrac{\ALG}{\OPT} \geq \cG$ always holds. 
\Halmos\endproof

\section{Proof of Theorem~\ref{thm:0.4324example}}
\label{sec:thm:0.4324example:proof}

\proof{Proof of Theorem~\ref{thm:0.4324example}.}
Let $u(\cdot)$ be a continuous function that describes the density in $[1-\qG, 1]$.
Let the random arrival sequence be $S$:
\begin{equation}
S = \left\{
\begin{aligned}
& (\underbrace{\eps, \eps, ..., \eps}_{(1-q)/\eps + 1 \text{ many}}, q), && \text{where } q \in [1-\qG, 1] \text{ conforms to } u(\cdot); \\
& (\qG, 1-\qG+\eps, 1-\qG+2\eps, ..., 1), && \text{with prob. } x; \\
& (\underbrace{\eps, \eps, ..., \eps}_{(1-\qG)/\eps +1 \text{ many}}, \qG), && \text{with prob. } y; \\
& (\eps, 1), && \text{with prob. } z;
\end{aligned}
\right.
\end{equation}
where
\begin{align*}
&x=\frac{1-2\cG}{1-2\qG} \approx 0.3711;& &y=\frac{1-2\cG}{\qG} \approx 0.4231;& &z=\cG-x \approx 0.0613;& &u(q)=\frac{x}{q}.&
\end{align*}

We can verify that $$x+y+z+\int_{1-\qG}^1 \frac{x}{q} \ \mathrm{d}q = x +  \frac{1-2\cG}{\qG} + (\cG - x) - \frac{1-2\cG}{1-2\qG} \cdot \ln{(1-\qG)} = 1,$$ by plugging the expressions into the equation and using $H(\cG,\qG)=0$ from \eqref{eqn:H=0}.
This equation shows that our construction conforms to a legitimate probability measure.

Following each realization of $S$, $\OPT(S)=1$. So we have $\bE_S[\OPT(S)] = 1$.

For any $\ALG=\THR(\tau),\tau\in[0,1]$, we enumerate all the potential values of $\tau$ in the following.\\
\textbf{Case 1}: $0 \leq \tau \leq \eps$. In this case,
\begin{align*}
\bE_S[\ALG(S)] & = \qG \cdot x + (1-\qG+\eps) \cdot y + \eps \cdot z + \int_{1-\qG}^1 (1-q+\eps) \cdot u(q) \mathrm{d}q \\
& = \qG \cdot x + (1-\qG) \cdot \frac{1-2\cG}{\qG} - \frac{1-2\cG}{1-2\qG} \cdot \ln{(1-\qG)} - x \cdot \qG + \eps \cdot (1-x)\\
& = \frac{1-2\cG}{\qG} - \frac{1-2\cG}{1-2\qG} \cdot \ln{(1-\qG)} - (1-2\cG) + \eps \cdot (1-x)\\
& = \cG + \eps \cdot (1-x),
\end{align*}
where the coefficient of $(1-x)$ next to the $\epsilon$ term is simplified using expression \eqref{eqn:H=0}; and the last equality is due to \eqref{eqn:H=0}.\\
\textbf{Case 2}: $\eps < \tau \leq \qG$. In this case,
\begin{align*}
\bE_S[\ALG(S)] & = \qG \cdot x + \qG \cdot y + 1 \cdot z + \int_{1-\qG}^1 q \cdot u(q) \mathrm{d}q \\
& = \qG \cdot \frac{1-2\cG}{1-2\qG} + \qG \cdot \frac{1-2\cG}{\qG} + \cG - \frac{1-2\cG}{1-2\qG} + \qG \cdot \frac{1-2\cG}{1-2\qG}\\
& = \cG + (1-2\cG)\left( 2\frac{\qG}{1-2\qG} + 1 - \frac{1}{1-2\qG}\right)\\
& = \cG
\end{align*}
\textbf{Case 3}: $\qG < \tau \leq 1 - \qG + \eps$. In this case,
\begin{align*}
\bE_S[\ALG(S)] & = (1 - \qG + \eps) \cdot x + 1 \cdot z + \int_{1-\qG}^1 q \cdot u(q) \mathrm{d}q \\
& = (1 - \qG) \cdot \frac{1-2\cG}{1-2\qG} + \cG - \frac{1-2\cG}{1-2\qG} + \qG \cdot \frac{1-2\cG}{1-2\qG} + \eps \cdot x\\
& = \cG + \eps \cdot x 
\end{align*}
\textbf{Case 4}: $1-\qG+\eps < \tau \leq 1$. In this case,
\begin{align*}
\bE_S[\ALG(S)] & \geq \tau \cdot x + 1 \cdot z + \int_{\tau}^1 q \cdot u(q) \mathrm{d}q \\
& = \tau \cdot x + \cG - x + (1-\tau) \cdot x\\
& = \cG
\end{align*}

In all, we have enumerated all the values that a threshold can take.
In all cases, the performance of the threshold $\THR(\tau)$ policy has an expected performance of no more than $\cG+ \eps \cdot (1-x)$.
But $\bE_S[\OPT(S)] = 1$.
By taking $\eps \to 0^+$ we finish the proof.
\Halmos\endproof

\section{Proofs of Theorems~\ref{thm:dependentSize} and~\ref{thm:generalization}}
\label{sec:app:dependentSize}

Theorem~\ref{thm:generalization} is a generalization of Theorem~\ref{thm:dependentSize}.
However, for better exposition, we prove both Theorems~\ref{thm:dependentSize} and~\ref{thm:generalization} separately.
We start with Theorem~\ref{thm:dependentSize}.

\proof{Proof of Theorem~\ref{thm:dependentSize}}
For any knapsack $j$, let $I_j$ be the set of items routed to it in Step 2 of Definition~\ref{defn:dependentSize} ($I_j$ includes items that are later discarded by the threshold of knapsack $j$).
Note that $I_j$ does not depend on the adoption of single-knapsack algorithms from Step 3.

Denote $\OPTp_{j} = \min\{\sum_{t \in I_j}s_{tj},B_j\}, \forall j \in [N]$.
Note that $\ALG_\mathsf{AW} = \sum_{j \in [N]} \OPTp_{j}$, due to the allowance of truncation in $\ALG_\mathsf{AW}$.

From Definition~\ref{defn:3/7alg}, we earn at least $\frac{3}{7} \cdot \OPTp_{j}$ from knapsack $j$ in expectation.
So when we focus on all knapsacks,
\begin{align*}
\ALG \geq \frac{3}{7} \sum_{j\in[N]} \OPTp_{j} \geq \frac{3}{7} (\frac{1}{2} \OPT_\mathsf{AW}(S)) \geq \frac{3}{14} \OPT,
\end{align*}
where the first inequality is because on each knapsack $\ALG$ earns at least $\frac{3}{7}$ fraction of what $\ALG_\mathsf{AW}$ does;
the second inequality is from Proposition~\ref{prop:AdWordsGreedy};
and the third inequality is simply the fact that $\OPT_\mathsf{AW} \geq \OPT$, because any optimal assignment when truncation is not allowed is a feasible solution to the problem when truncation is allowed.
\Halmos\endproof

Following Theorem~\ref{thm:dependentSize}, we generalize Definition~\ref{defn:dependentSize} and present Definition~\ref{defn:generalization} below.
Let $\ALG_{\mathsf{AW}}$ be any algorithm designed for the AdWords problem;
let $\ALG_{\mathsf{SK}}$ be any algorithm designed for the single knapsack problem.
Similar to Definition~\ref{defn:dependentSize}, we give a formal statement of combining the above two algorithms as follows.

\begin{definition}
\label{defn:generalization}
Let $d_{tj}$ denote the ``phantom'' capacity filled in each knapsack $j \in [N]$ at time $t\in[T]$, if all the items routed to knapsack $j$ were accepted, with truncation allowed. The algorithm proceeds as follows:
\begin{enumerate}
\item Initialization: at time $t=0$, set $b_{1j} = 0, d_{1j} = 0, \forall j \in [N]$. 
\item For each item $t$, use $\ALG_{\mathsf{AW}}$ to route item $t$ to knapsack $\tilde{j}$, and update (i) $d_{(t+1) \tilde{j}} = d_{t \tilde{j}} + s_{t \tilde{j}}$; (ii) $d_{(t+1) j} = d_{t j}, \forall j \ne \tilde{j}$.
\item For item $t$ that is routed to knapsack $\tilde{j}$, use $\ALG_{\mathsf{SK}}$ to accept or discard item $t$. If item $t$ is accepted, update (i) $b_{(t+1) \tilde{j}} = b_{t \tilde{j}} + s_{t \tilde{j}}$; (ii) $b_{(t+1) j} = b_{t j}, \forall j \ne \tilde{j}$; if item $t$ is discarded, $b_{(t+1)j} = b_{tj}, \forall j \in [N]$.
\end{enumerate}
\end{definition}


\proof{Proof of Theorem~\ref{thm:generalization}}
For any knapsack $j$, let $I_j$ be the set of items routed to it in Step 2 of Definition~\ref{defn:generalization} ($I_j$ includes items that are later discarded by $\ALG_{\mathsf{SK}}$).
Note that $I_j$ does not depend on the adoption of single-knapsack algorithms from Step 3.

Denote $\OPTp_{j} = \min\{\sum_{t \in I_j}s_{tj},B_j\}, \forall j \in [N]$.
Note that $\ALG_\mathsf{AW} = \sum_{j \in [N]} \OPTp_{j}$, due to the allowance of truncation in $\ALG_\mathsf{AW}$.

Note that $\ALG_{\mathsf{SK}}$ is $\Gamma_{\mathsf{SK}}$-competitive for the single knapsack problem relative to the optimal fractional packing.
We earn at least $\Gamma_{\mathsf{SK}} \cdot \OPTp_{j}$ from knapsack $j$ in expectation.
So when we focus on all knapsacks,
\begin{align*}
\ALG \geq \Gamma_{\mathsf{SK}} \sum_{j\in[N]} \OPTp_{j} \geq \Gamma_{\mathsf{SK}} (\Gamma_{\mathsf{AW}} \OPT_\mathsf{AW}(S)) \geq \Gamma_{\mathsf{AW}}\Gamma_{\mathsf{SK}} \OPT,
\end{align*}
where the first inequality is because on each knapsack $\ALG$ earns at least $\Gamma_{\mathsf{SK}}$ fraction of what $\ALG_\mathsf{AW}$ does;
the second inequality is from Proposition~\ref{prop:AdWordsGreedy};
and the third inequality is simply the fact that $\OPT_\mathsf{AW} \geq \OPT$, because any optimal assignment when truncation is not allowed is a feasible solution to the problem when truncation is allowed.
\Halmos\endproof

\section{Proof of Theorem~\ref{thm:knapsackDEexample}}
\label{sec:app:knapsackDEexample}

\proof{Proof of Theorem~\ref{thm:knapsackDEexample}.}
Consider the example we described above, where $N=4$, and $\alpha=1 - \frac{12}{7} \eps$.
By equation~\eqref{eqn::upperTriangular} above,
$$\bE_S[\OPT(S)]= (1-\frac{12}{7}\eps) N \eps + (\frac{12}{7}\eps)N = \frac{76}{7} \eps - \frac{48}{7} \eps^2.$$

Now we analyze the maximum possible value of $\bE_S[\ALG(S)]$.  As discussed before, any algorithm is characterized by $e\in\{0,1,\ldots,4\}$, which is the number of size-$\eps$ items accepted.

\textbf{Case 1}: $e=0$.
With probability $\alpha$, the arrival sequence terminates with $0$ accepted;
with probability $1-\alpha$, there are $N$ more items.
We enumerate all the $24$ possibilities, to find there are $6$ cases that a deterministic algorithm accepts $2$ of them, $17$ cases that accepts $3$, and $1$ case that accepts $4$.
In expectation we fill $67/24$ into the knapsacks.
In this case $$\frac{\bE_S[\ALG(S)]}{\bE_S[\OPT(S)]} = \frac{\frac{12}{7} \cdot \eps \cdot \frac{67}{24}}{\frac{76}{7} \cdot \eps - \frac{48}{7} \cdot \eps^2} = \frac{67}{152 - 96\eps}.$$

\textbf{Case 2}: $e=1$.
With probability $\alpha$, the arrival sequence terminates with $\eps$ accepted;
with probability $1-\alpha$, there are $N$ more items.
Out of all the $24$ possibilities, there are $16$ cases that a deterministic algorithm accepts $2$ of them, and $8$ cases that accepts $3$.
In expectation we fill $56/24$ into the knapsacks.
In this case $$\frac{\bE_S[\ALG(S)]}{\bE_S[\OPT(S)]} = \frac{1 \cdot \eps + \frac{12}{7} \cdot \eps \cdot \frac{56}{24}}{\frac{76}{7} \cdot \eps - \frac{48}{7} \cdot \eps^2} = \frac{35}{76-48\eps}.$$

\textbf{Case 3}: $e=2$.
With probability $\alpha$, the arrival sequence terminates with $2 \eps$ accepted;
with probability $1-\alpha$, there are $N$ more items.
Out of all the $24$ possibilities, there are $6$ cases that a deterministic algorithm accepts $1$ of them, and $18$ cases that accepts $2$.
In expectation we fill $42/24$ into the knapsacks.
In this case $$\frac{\bE_S[\ALG(S)]}{\bE_S[\OPT(S)]} = \frac{2 \cdot \eps + \frac{12}{7} \cdot \eps \cdot \frac{42}{24}}{\frac{76}{7} \cdot \eps - \frac{48}{7} \cdot \eps^2} = \frac{35}{76-48\eps}.$$

\textbf{Case 4}: $e=3$.
With probability $\alpha$, the arrival sequence terminates with $3 \eps$ accepted;
with probability $1-\alpha$, there are $N$ more items.
The algorithm must be able to fill one item into the unfilled knapsack in the first round of phase two.
In this case $$\frac{\bE_S[\ALG(S)]}{\bE_S[\OPT(S)]} = \frac{3 \cdot \eps + \frac{12}{7} \cdot \eps \cdot 1}{\frac{76}{7} \cdot \eps - \frac{48}{7} \cdot \eps^2} = \frac{33}{76-48\eps}.$$

\textbf{Case 5}: $e=4$.
With probability $\alpha$, the arrival sequence terminates with $4 \eps$ accepted;
with probability $1-\alpha$, there are $N$ more items.
But the algorithm cannot fill in any because all the knapsacks are all occupied with $\eps$'s.
In this case $$\frac{\bE_S[\ALG(S)]}{\bE_S[\OPT(S)]} = \frac{4 \cdot \eps + \frac{12}{7} \cdot \eps \cdot 0}{\frac{76}{7} \cdot \eps - \frac{48}{7} \cdot \eps^2} = \frac{28}{76-48\eps}.$$

In all cases, any policy has an expected performance of no more than $\frac{35}{76 - 48 \eps}$.
By taking $\eps \to 0^+$ we finish the proof.
\Halmos\endproof

\section{Additional Simulations}
\label{sec:AdditionalSimu}

Recall that in Section~\ref{sec:Techniques} we discussed the intuition that we should choose a CDF that lives at the intersection.
Different CDFs that live at the intersection will have different performance.
In this section, we compare two different CDFs through simulations.

Similar to the simulation setup as in Section~\ref{sec::computational}, we consider the arrival sequence as defined in \eqref{eqn:example}.
We compute the expected revenue from our proposed random threshold algorithm from Definition~\ref{defn:3/7alg}, named \textbf{Random-Threshold}; the (deterministic) revenue from first-come-first-serve policy, named \textbf{FCFS}; and the expected revenue from the algorithm suggested in \citet{han2015randomized}, named \textbf{HKM15}.
Additionally, we compute the expected revenue from a random threshold algorithm whose CDF is specified as
\begin{align*}
F(x) = \left\{
\begin{aligned}
& \frac{\frac{4}{7}-x}{1-2x}, & \quad x \in [0, 1/3]\\
& \frac{8}{7}-\frac{1}{7x}, & \quad x \in (1/3,1]
\end{aligned}
\right.
\end{align*}
We refer to this algorithm as \textbf{Random-Threshold 2} in our simulation. 

\begin{figure}[!tb]
\centering
\includegraphics[width=0.95\linewidth]{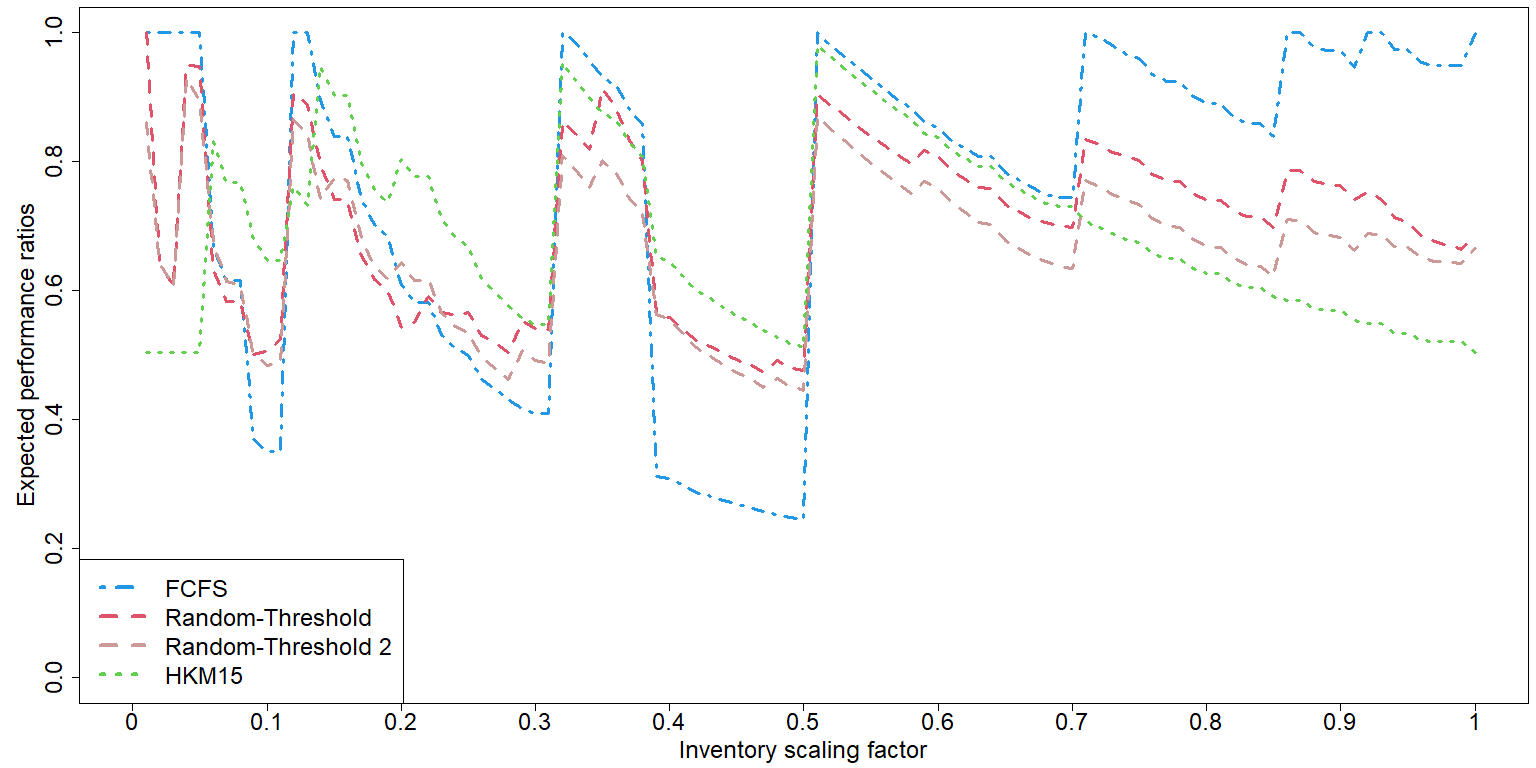} 
\caption{Computational performance on one real arrival sequence shown in Equation~\eqref{eqn:example}.}
\label{fig:additional}
\end{figure}

The results are shown in Figure~\ref{fig:additional}, where we have divided all the numbers by its corresponding offline optimal integer packing.
The offline optimal packing serves as an upper bound, so that the performance ratio is always between $0$ and $1$, with higher ratios indicating better performance.
As we can see in Figure~\ref{fig:additional}, although \textbf{Random-Threshold} and \textbf{Random-Threshold 2} have two CDFs that live at the same intersection, they have different performance.
\textbf{Random-Threshold} slightly outperforms \textbf{Random-Threshold 2}.

\end{document}